\documentclass{article} % For LaTeX2e
\usepackage{iclr2024_conference,times}

\usepackage{amsmath,amsfonts,bm}

\def\eqref#1{equation~\ref{#1}}
\def\1{\bm{1}}

\def\rv{{\textnormal{v}}}

\def\rvb{{\mathbf{b}}}

\def\rve{{\mathbf{e}}}

\def\rvh{{\mathbf{h}}}

\def\rvm{{\mathbf{m}}}

\def\rvv{{\mathbf{v}}}

\def\rvx{{\mathbf{x}}}

\def\rmI{{\mathbf{I}}}

\def\vzero{{\bm{0}}}

\def\vmu{{\bm{\mu}}}

\def\va{{\bm{a}}}
\def\vb{{\bm{b}}}
\def\vc{{\bm{c}}}

\def\vv{{\bm{v}}}

\def\vx{{\bm{x}}}

\def\mA{{\bm{A}}}

\def\mP{{\bm{P}}}

\DeclareMathAlphabet{\mathsfit}{\encodingdefault}{\sfdefault}{m}{sl}
\SetMathAlphabet{\mathsfit}{bold}{\encodingdefault}{\sfdefault}{bx}{n}

\def\gA{{\mathcal{A}}}

\def\gC{{\mathcal{C}}}

\def\gG{{\mathcal{G}}}

\def\gM{{\mathcal{M}}}
\def\gN{{\mathcal{N}}}

\def\gP{{\mathcal{P}}}

\def\gS{{\mathcal{S}}}

\newcommand{\R}{\mathbb{R}}

\usepackage{hyperref}
\usepackage{url}
\usepackage{xspace} % used in \method

\usepackage{url}            % simple URL typesetting
\usepackage{booktabs}       % professional-quality tables
\usepackage{amsfonts}       % blackboard math symbols
\usepackage{nicefrac}       % compact symbols for 1/2, etc.
\usepackage{microtype}      % microtypography
\usepackage{xcolor}         % colors

\usepackage{amsmath}
\usepackage{amssymb}
\usepackage{mathtools}
\usepackage{amsthm}
\usepackage{dsfont}
\usepackage{diagbox}
\usepackage{adjustbox}
\usepackage{multirow}
\usepackage{makecell}
\usepackage{wrapfig}
\usepackage{subfigure}
\usepackage{graphicx}

\usepackage{caption}
\usepackage{algorithm}
\usepackage{algpseudocode}
\usepackage{tabularx}

\usepackage{paralist} % for compactitem

\usepackage{yhmath}
\usepackage{pifont}
\usepackage{float}
\usepackage{array}
\usepackage{enumitem}
\usepackage[utf8]{inputenc}

\theoremstyle{plain}

\theoremstyle{definition}

\theoremstyle{remark}

\usepackage[capitalize,noabbrev]{cleveref}
\usepackage[textsize=tiny]{todonotes}
\crefname{section}{Section}{Sections}
\Crefname{section}{Section}{Sections}
\Crefname{table}{Table}{Tables}
\crefname{table}{Table}{Tables}

\usepackage[compact]{titlesec}
\usepackage{sidecap}
\titlespacing*{\section}{0pt}{*0.34}{*0.34}
\titlespacing*{\subsection}{0pt}{*0.34}{*0.34}
\titlespacing*{\subsubsection}{0pt}{*0.34}{*0.34}

\allowdisplaybreaks

\title{Controllable and Decomposed Diffusion Models for Structure-based Molecular Optimization}

\author{
Xiangxin Zhou$^{1,2,3}$\thanks{Equal Contribution. Work was done during Xiangxin's and Xiwei's internship at ByteDance.}\,\,  Xiwei Cheng$^{3,4*}$ Yuwei Yang$^3$ Yu Bao$^3$ Liang Wang$^{1,2}$ Quanquan Gu$^{3}$\thanks{Corresponding Author: Quanquan Gu (quanquan.gu@bytedance.com).}\\
$^1$School of Artificial Intelligence, University of Chinese Academy of Sciences
\\
$^2$Center for Research on Intelligent Perception and Computing (CRIPAC), \\
\, State Key Laboratory of Multimodal Artificial Intelligence Systems (MAIS), \\
\, Institute of Automation, Chinese Academy of Sciences (CASIA)
\\
$^3$ByteDance Research
\\
$^4$Halıcıoğlu Data Science Institute, University of California San Diego 
}

\newcommand{\method}{\textsc{DecompOpt}\xspace}

\newcommand{\revise}[1]{{{\textcolor{black}{#1}}}}

\iclrfinalcopy % Uncomment for camera-ready version, but NOT for submission.
\begin{document}

\maketitle

\begin{abstract}

Recently, 
3D generative models have shown promising performances in structure-based drug design by learning to generate ligands given target binding sites.
However, only modeling the target-ligand distribution can hardly fulfill one of the main goals in drug discovery -- designing novel ligands with desired properties, e.g., high binding affinity, easily synthesizable, etc. This challenge becomes particularly 
pronounced when the target-ligand pairs used for training do not align with these desired properties.
%especially, if the target-ligand pairs used for training do not satisfy the requirements as well. 
Moreover, most existing methods aim at solving \textit{de novo} design task, while many generative scenarios requiring flexible controllability, such as R-group optimization and scaffold hopping, have received little attention. 
In this work, we propose \method, a structure-based molecular optimization method based on a controllable and decomposed diffusion model. 
\method presents a new generation paradigm which combines optimization with conditional diffusion models to achieve desired properties while adhering to the molecular grammar,
Additionally, 
\method offers a unified framework covering both \textit{de novo} design and controllable generation. 
To achieve so, ligands are decomposed into substructures which allows fine-grained control and local optimization. 
Experiments show that \method can efficiently generate molecules with improved properties than strong \textit{de novo} baselines, and demonstrate great potential in controllable generation tasks.
\end{abstract}
\section{Introduction}
\begin{wrapfigure}{r}{0.46\textwidth}
\vspace{-0.1in}
\centering
    \includegraphics[width=0.45\textwidth]{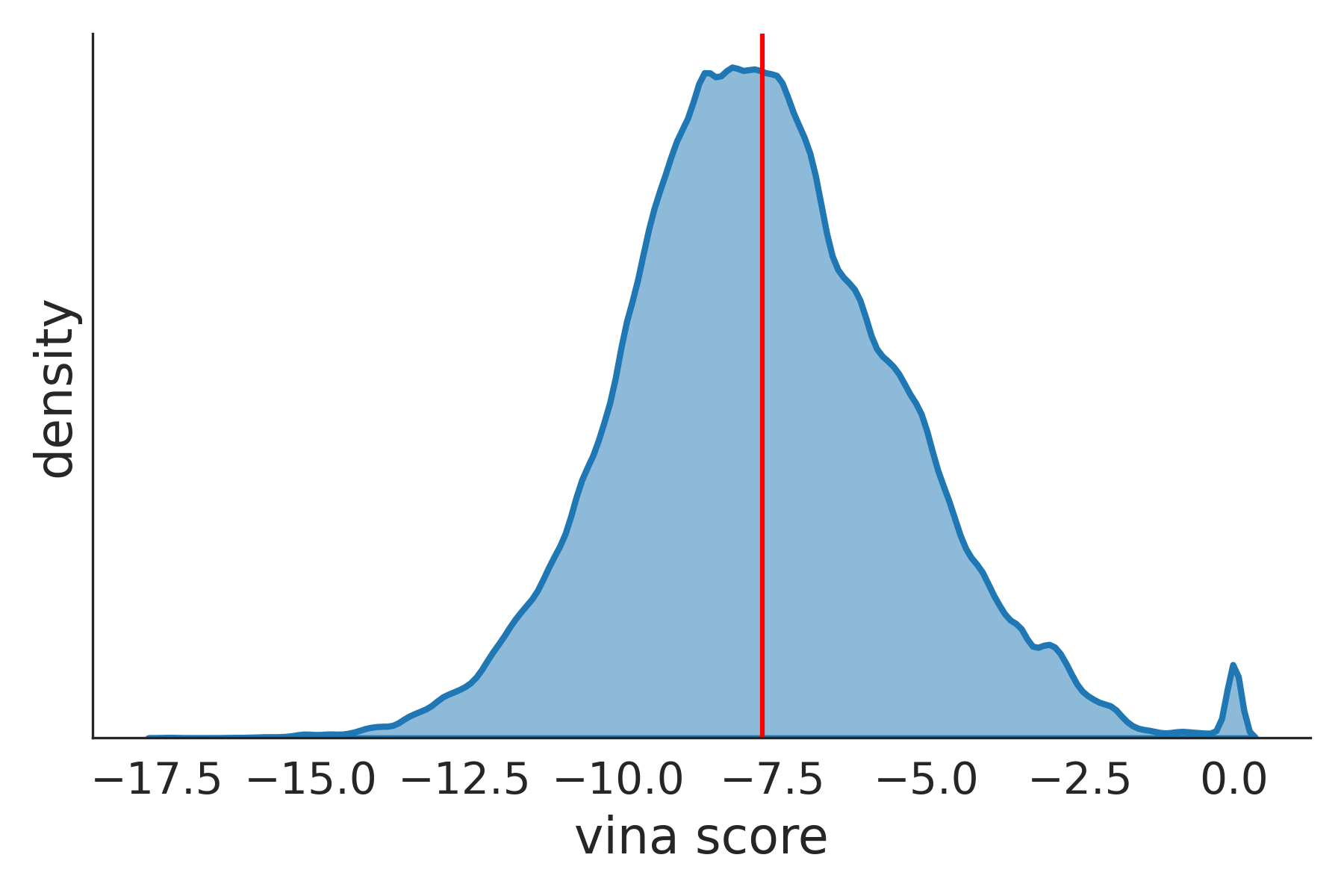}
    \vspace{-0.2cm}
    \captionof{figure}{Vina Scores distribution of protein-ligand pairs in CrossDocked2020 dataset. $-8.18$ kcal/mol, marked by the red vertical line, is a commonly used value representing moderate binding affinity.}
    \label{fig:motivation}
    \vspace{-4mm}
\end{wrapfigure}

Structure-based drug design (SBDD) \citep{anderson2003process} is an approach that involves designing drug molecules based on the 3D structure of a target. 
The goal of SBDD is to generate ligands with desired properties which can bind tightly to the target binding site. 
Recently, several works have cast SBDD into a conditional 3D molecular generation task and achieved remarkable success thanks to the powerful deep generative models. In these models, the target binding site serves as the condition and the conditional distribution of ligands is learnt in a data-driven manner using various generative models.
\citet{peng2022pocket2mol} and \citet{zhang2022molecule} proposed to generate molecules given pockets in an auto-regressive fasion using atoms or fragments as building blocks respectively. \citet{guan3d,schneuing2022structure,lin2022diffbp} use diffusion models to generate ligands by modeling atom types and positions.

Generative models are powerful approaches for extracting the underlying molecular grammar (e.g., the reasonable atomic valence, stable molecular conformation, etc.). However, they cannot generate molecules with desired properties if the training data do not align with these properties as well. Indeed, unsatisfying data quality is a common challenges in drug discovery \citep{vamathevan2019applications}. As \cref{fig:motivation} shows, the ligands in CrossDocked2020 \citep{francoeur2020three}, a widely used training dataset for SBDD models, have moderate binding affinities measured by molecular docking scores. Solely maximizing the likelihood of training data can mislead SBDD models and cause inefficiency in generating potential drug candidates. To overcome this limitation, molecular optimization \citep{xie2021mars,fu2022reinforced} offers a direct path for searching molecules with desired properties in the broad chemical space. However, its application to 3D molecule generation remains unexplored.   

%However, the evaluation metrics (e.g., QED, SA, and Vina Score) measure the properties of the generated ligand molecules.  Recently, \citet{pmlr-v202-guan23a,zhang2023learning} proposed to incorporate more domain knowledge into generative models for SBDD, which may encourage the model to generate desired ligand molecules. These generative models mimic realistic molecules well in many aspects, such as topology, 3D structures, and molecular properties, among others. Nevertheless, a frustrating fact is that the pocket-ligand pair data used as training data are not satisfying, especially in terms of some molecular properties. As \cref{fig:motivation} shows, the ligand molecules in the training set of CrossDocked2020 \citep{francoeur2020three} are far from drug-like. Only maximizing the likelihood of training data misleads the SBDD models and causes inefficiency in generating potential drug candidates. This limitation of the generative perspective motivates us to introduce property optimization into the generative models. 

 On the other hand, current SBDD models are mostly limited to \textit{De novo} design \citep{hartenfeller2011novo}, which focuses on generating ligands from scratch, is the main task that most efforts have been devoted to. 
However, controllable molecular generation
% \footnote{\revise{Note that rollable molecular generation is referred to as generating molecules with reference structures or substructures in this paper.}}
scenarios, such as R-group design \citep{takeuchi2021r} (also known as biosteric replacement) and scaffold hopping \citep{bohm2004scaffold}, are equally, if not more, important. Unlike \textit{de novo} design, controllable generation tasks start from an existing compound, and only modify a local substructure to improve the synthetic accessibility, potency and drug-likeness properties or to move into novel chemical space for patenting \citep{langdon2010bioisosteric}. Controllable generation aims to utilize prior knowledge, such as a known active compound, in the design process to increase the chance of finding promising candidates. 
Some initial efforts have been made to address controllable molecular generation problem. For example, \citet{igashov2022equivariant, imrie2020deep, imrie2021deep, huang20223dlinker} propose to use generative models to design linkers between the given fragments. However, these methods are designed for a specific controllable generation task and cannot be generalized.
%However, the invention of a drug molecule often goes through several rounds of structural modification and experimental validation in order to achieve optimal potency and biological activity profile. 
%More specifically, medicinal chemists usually use analog-based drug design to derive a series of molecules, which share a common core structure and only differs by one functional group, to improve local properties and investigate structure-activity relation. \citep{fischer2010analogue}
%Therefore, controllable generation, which only designs or modifies a local substructure, is equally important if not more. 
%\xiangxin{e.g., DiffLinker. Not general for all downstream tasks.}
%Another discrepancy between research community and industry of drug discovery is that machine learning researchers seek to propose simple but effective methods while industry expects flexible molecule design methods that can satisfy various specific requirements proposed by pharmacists and chemists. Generally, domain experts prefer models with a certain degree of controllability so that they can easily incorporate their domain knowledge into ML-assisted molecule design. \xiangxin{Need Yuwei's help to introduce some well-recognized requirements here such as R-group optimization and need to be polished}

%To bridge the aforementioned gap between machine learning research and pharmaceutical practice, 
To overcome the aforementioned challenges and limitations in existing SBDD approaches, we propose \method, a controllable and decomposed diffusion model for structure-based molecular optimization. 
\method combines diffusion models with optimization algorithm to harness the advantages of both approaches. Diffusion models are used to extract molecular grammar in a data-driven fashion, while optimization algorithm is used to effectively optimize the desired properties. 
Furthermore, \method offers a unified generation framework for both \textit{de novo} design and controllable generation through ligands decomposition.
Notably, a ligand that binds to a target binding site can be naturally decomposed into several substructures, i.e., arms and scaffold, where arms locally interact with corresponding \underline{sub}pockets and scaffold links all arms to form a complete molecule. Such decomposition motivates us to design a conditional diffusion models in the decomposed drug space which ensures flexible and fine-grained control over each substructure. 
%\xiangxin{adjust the order: first mention control, second mention decomposition} 
%Based on this model, we further propose an optimization process that can gradually promote the properties of generated ligand molecules by discovering and utilizing higher-quality substructures.
%\xiangxin{mention FBDD}
%The idea is similar to fragment-based drug discovery (FBDD) \citep{murray2009rise}. 
%Introducing optimization into generation better aligns with the goal of drug discovery while also keeping considerable diversity. This new paradigm combines the advantages of both optimization and generation and provides a new perspective for structure-based drug design. 
We highlight our main contributions as follows:
\begin{itemize}[leftmargin=*]
\item We propose a new molecular generation paradigm, which combines diffusion models with iterative optimization to learn molecular grammar and optimize desired properties simultaneously.
\item We design a unified generation framework for \textit{de novo} design and controllable generation via a controllable and decomposed diffusion model.
%\item Based on the model, we propose a structure-based molecular optimization process that can gradually discover more desired substructures and improve the properties of generated ligands conditioned on them. \xiangxin{A new paradigm, better align with xxx}
\item For \textit{de novo} design, our method can generate ligands with an average Vina Dock score of $-8.98$ and a Success Rate of $52.5\%$,
% better binding affinities with a -8.39 Vina Dock score and a 24.46\% higher success rate, 
achieving a new SOTA on the CrossDocked2020 benchmark.
%, while keeping considerable diversity.
\item For controllable generation, our method shows promising results in various practical tasks of SBDD, including R-group design and scaffold hopping.
\end{itemize}

\section{Related Work}
\paragraph{Molecule Generation} 
Deep generative models have shown promising results in molecule generation. In the last decade, researchers have explored various representations and models for molecule generation. Molecules can be represented in 1D (e.g., SMILES \citep{weininger1988smiles}, SELFIES \citep{krenn2020self}), 2D (i.e., molecular graph \citep{bonchev1991chemical}), and 3D. Among them, 3D representations attract recent attention since they capture the complete information of molecules, and have better potential to generate and optimize molecules with regard to 3D properties, such as bioactivity for a given target \citep{baillif2023deep}.
%Spatial information is essential to model some molecular properties and functions. 
%Thus, recent works \citep{gebauer2019symmetry,hoogeboom2022equivariant} focus on directly generating 3D molecules in 3D space. 

SBDD represents an important application for 3D molecule generation. 
%One of the most important applications of molecule generation is structure-based drug design \citep{anderson2003process}, which aims to generated ligand molecules given a protein pocket. 
%\citet{skalic2019target,xu2021novo} generate SMILES of ligands based on protein contexts. \citet{tan2022target} proposed to use flow-based models to generate molecular graphs given sequences of target proteins. 
\citet{ragoza2022generating} generate 3D molecules in atomic density grids using a variational autoencoder \citep{kingma2013auto}.
\citet{luo20213d, liu2022generating,peng2022pocket2mol} propose to generated atoms (and bonds) auto-regressively in 3D space, while \citet{zhang2022molecule} use fragment as building blocks instead. \citet{guan3d,lin2022diffbp,schneuing2022structure} introduce SE(3)-equivariant diffusion models for SBDD. 
More recent works have incorporated domain knowledge into 3D generative models, such as the correspondence between local fragments and subpockets. \citet{guan3d} suggest to break ligands into substructures and model them using decomposed priors in a diffusion framework, leading to remarkably improved binding affinities of the generated molecules.  
\citet{zhang2023learning} propose a subpocket prototype-augmented 3D molecule generation scheme to establish the relation between subpockets and their corresponding fragments. 
Existing methods based on deep generative models are powerful at distribution learning. However, when the training examples do not have the desired properties, these models can hardly generate out-of-distribution samples with these properties.

\noindent\textbf{Molecule Optimization}
Optimization-based algorithms are another popular approach to design drug molecules. Methods within this category rely on predefined computable objectives to guide the optimization. Various optimization methods have been proposed for 2D drug design. 
JTVAE \citep{jin2018junction} uses Bayesian optimization in the latent space
to indirectly optimize molecules. 
Reinforcement learning is used to manipulate SMILES strings \citep{olivecrona2017molecular} and 2D molecular graphs \citep{zhou2019optimization, jin2020multi}. 
MARS \citep{xie2021mars} leverages adaptive Markov chain Monte Carlo sampling to accelerate the exploration of chemical space. 
RetMol develops a retrieval-based generation scheme for iteratively improving molecular properties. 
Genetic algorithm is also a popular choice. 
GA+D \citep{Nigam2020Augmenting} uses deep learning enhanced genetic algorithm to design SELFIES strings. 
Graph-GA \citep{jensen2019graph} conducts genetic algorithm
on molecular graph representation. 
GEGL \citep{ahn2020guiding} adopts genetic algorithm to generate high quality samples for imitation learning by deep neural networks. AutoGrow 4 \citep{spiegel2020autogrow4} and RGA \citep{fu2022reinforced} are genetic algorithms for SBDD which incorporate target structures in molecular optimization. Both of them use molecular docking score as an objective to optimize the fitness between target structure and the generated ligands. In addition, RGA uses neural models to stablize genetic algorithm and includes target structures as a condition in its modeling. To our best knowledge, there are limited efforts on generating 3D molecules using molecular optimization. 

Although optimization algorithms offers a direct approach to achieve desired properties, they require computable and accurate objectives to guide the exploration. However, not all desired properties for drug design can be easily formulated as objectives, such as molecular validity. Considering the benefits of both generative models and optimization algorithm, it is reasonable to combine them to achieve further enhanced results.

\noindent\textbf{Controllable Generation} 
\textit{De novo} design aims to generate molecules from scratch, and the above-mentioned methods mainly focus on this task. 
Besides it, another line of research focuses on controllable molecule generation, which requires generating or optimizing partial molecules. R-group design is a task to decorate a fixed scaffold with fragments to enhance the desired properties. \citet{langevin2020scaffold} and \citet{maziarz2022learning,imrie2021deep} propose to constrain the scaffold region using SMILES-based and 2D graph-based models. However, similar attempts have been rarely observed in 3D molecule generation. 
Scaffold hopping, on the other hand, requires the replacement of the scaffold to explore novel chemical space while maintaining the favorable decorative substructures. \citet{imrie2020deep, imrie2021deep} propose autoregressive models to design 2D linker conditioning on geometric features of the input fragments and pharmacophores. \citet{huang20223dlinker, igashov2022equivariant} extend the application to the 3D space using variational autoencoders and diffusion models. However, existing 3D controllable generation methods are specifically designed for a single task. There lacks an unified framework to cover all possible conditional generation tasks, as well as \textit{de novo} design.
\section{Method}

\begin{figure*}[ht]
\centering
% \raggedleft
% \flushleft
% \vspace{-0.17in}
% \vspace{-0.07in}
\includegraphics[width=1.0\textwidth]{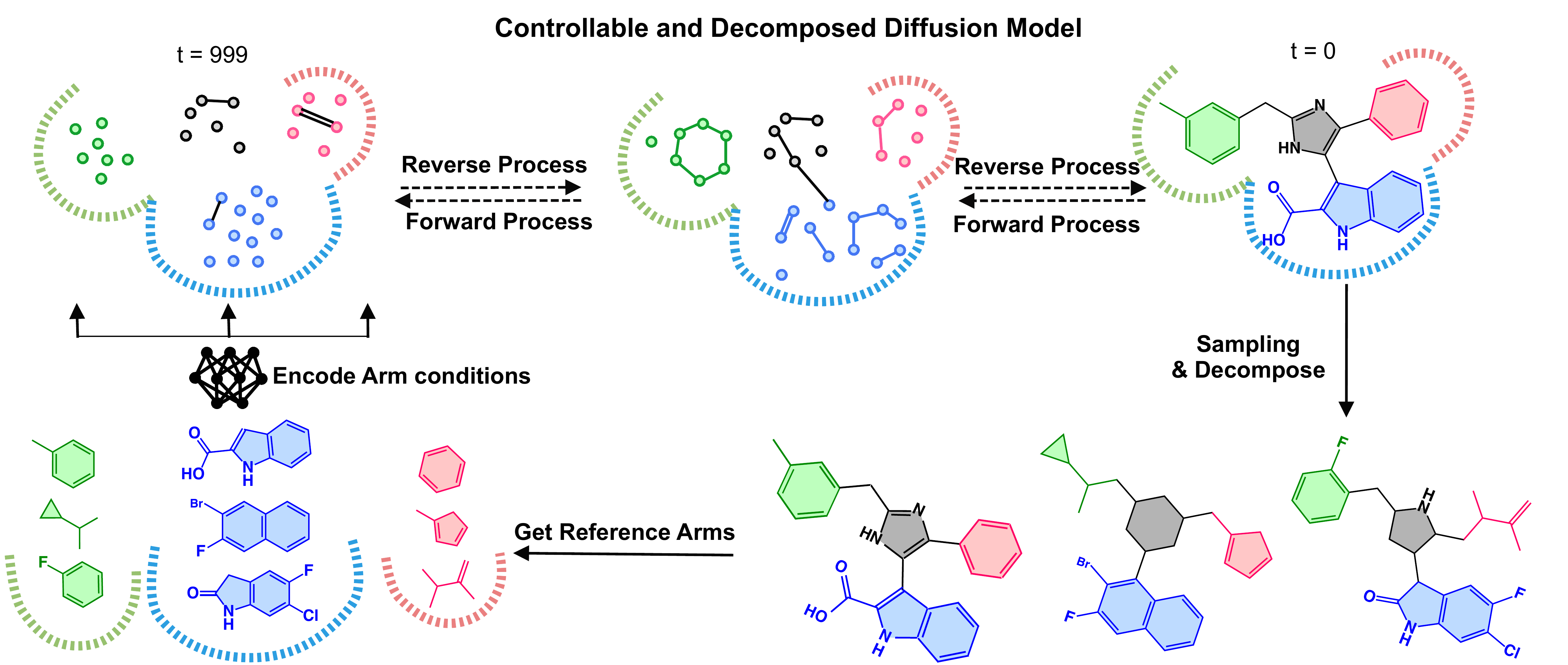}
    % \vspace{-0.1in}
  \caption{Illustration of \method. In each iteration of optimization: (1) For each subpocket, a reference arm is sampled from the ordered arm list. (2) The controllable and decomposed diffusion model generated ligand molecules based on arm (and subpocket) conditions. (3) The generated ligand molecules are collected and further decomposed into scaffolds and arms. (4) Poor arms in the ordered arm lists are replaced with the new arms that show better properties.}{\label{fig:framework}}
% \vspace{-0.23 in}
\end{figure*}

In this section, we will present our method, named \method, as illustrated in \cref{fig:framework}. In \cref{subsec:controllable_diffusion}, we show how to design a controllable and decomposed diffusion model that can generate ligand molecules conditioning on both protein subpockets and reference arms. In \cref{subsec:optimization_procedure}, we show how to efficiently optimize the properties of the generated ligand molecules in the decomposed drug space by improving the arm conditions. 

\subsection{Controllable and Decomposed Diffusion Models} 
\label{subsec:controllable_diffusion}

A ligand molecule that binds to a specific protein pocket can be naturally decomposed into several components (i.e., arms and scaffold). The arms of a ligand molecule locally interact with subpockets of the target protein. Notably, they are the main contributors to binding affinity. The scaffold links all the arms to form a complete molecule. Inspired by this, \citet{pmlr-v202-guan23a} introduced decomposed priors to diffusion models for SBDD. The decomposed priors not only induce a better variational lower bound as the training objective but also provides possibilities to achieve controllability in molecular generation. Specifically, the decomposed priors allow for relatively independent modeling of each arm. To combine generative models with optimization, a flexible and controllable generation framework is need. Thus we propose a controllable and decomposed diffusion model that allows for fine-grained control over the arms of the generated ligands. Considering the different functionalities of the arms and scaffold, we only control the arms that play important roles in interaction with pockets, and leave room for the generative model on the scaffold to achieve the trade-off between controllability and diversity. 

Provided with a target binding site that can be represented as $\gP=\{(\vx^\gP_i,\vv^\gP_i)\}_{i\in\{1,\ldots,N_\gP\}}$, we aim to generate a ligand molecule that can be represented as $\gM=\{(\vx^\gM_i,\vv^\gM_i,\vb^\gM_{ij})\}_{i,j\in\{1,\ldots,N_\gM\}}$. $N_\gP$ and $N_\gM$ are the number of atoms in the protein pocket and ligand molecule, respectively. 
Here $\vx\in\R^3$, $\vv\in\R^d$, $\vb\in\R^5$ denote the atom position, atom type, and bond type, respectively. 
A ligand molecule $\gM$ can be decomposed into a scaffold $\gS$ and several arms $\{\gA_{k}\}_{k\in\{1,\ldots,K\}}$, where $\gM = \gS \cup \gA_1 \cup \cdots \cup \gA_K$. 
We denote the subpocket that is within $10\text{\r{A}}$ of the atoms of the $k$-th arm $\gA_k$ as $\gP_k$.
The controllable and decomposed diffusion model is expected to generate a ligand molecule $\gM$ given a protein pocket $\gP$ and several reference arms $\{{\gA}_k\}$, which can be formulated as modeling the conditional distribution $q(\gM|\gP,\{{A}_k\})$. And the arms of the generated ligand molecules are expected to be similar to the corresponding reference arms. 

Generally, there are two critical modules in our model: a condition encoder and a diffusion-based decoder. We employ an SE(3)-equivariant neural network \citep{satorras2021n}, named EGNN, to encode a reference arm $\gA_k$ and its surrounding subpocket $\gP_k$. We introduce subpockets here to include information of intermolecular interaction and relative positions. Specifically, we build a $k$-nearest-neighbor (knn) geometric graph on the complex of the reference arm and its surrounding subpocket and apply the EGNN to learn its representation as follows: $[ \mA_k, \mP_k ]=\text{Enc}(\gA_k, \gP_k)$, where $\mA_k\in\R^{|\gA_k|\times D}$, $\mP_k\in\R^{|\gP_k|\times D}$, $[ \cdot ]$ denotes concatenation along the first dimension. Each row of $\mA_k$ (resp. $\mP_k$) corresponds to a condition feature of an atom in the reference arm $\gA_k$ (resp. the subpocket $\gP_k$). $\va_k=\text{Agg}([ \mA_k, \mP_k ]) \in \R^D$ is the global SE(3)-invariant condition feature aggregated from the atom-wise condition features. 

The diffusion model first perturbs samples by iteratively injecting random noises which are independent of the arm conditions. This leads to the forward process as follows: 
\begin{align} \label{eq:forward_process}
    q(\gM_{1:T}|\gM_0,\gP,\{{\gA}_k\}) = q(\gM_{1:T}|\gM_0,\gP) = \prod_{t=1}^T q(\gM_t|\gM_{t-1},\gP),
\end{align}
where $\gM_0\sim q(\gM|\gP,\{\gA_k\})$ and $\{\gM_t\}_{t=1}^T$ is a sequence of perturbed ligands.
We will introduced the aforementioned condition features into the reverse (generative) process of the diffusion model as:
\begin{align} \label{eq:reverse_process}
    p_\theta(\gM_{0:T-1}|\gM_T,\gP,\{{\gA}_k\}) = \prod_{t=1}^T p_\theta(\gM_{t-1}|\gM_t,\gP,\{{\gA}_k\}).
\end{align}
To model $p_\theta(\gM_{t-1}|\gM_t,\gP,\{{\gA}_k\})$, for the input (i.e., the ligand molecule being generated at time step $t$) of the diffusion-based decoder, we denote the SE(3)-invariant feature of its each arm atom as $\vv^{\gA}_i$ and each scaffold atom as $\vv^{\gM}_i$. For each arm atom that belongs to its $k$-th arm, we incorporate the aforementioned arm condition as $\Tilde{\vv}^\gA_i=\text{MLP}([ \vv^{\gA}_i, \va_k ])$. For each scaffold atom, we do not introduce any condition (i.e., $\Tilde{\vv}^\gS_i\coloneqq \text{MLP}(\vv^\gS_i)$) and leave enough room for the generative model to generate diverse scaffolds. For each atom of the protein pocket, we let $\vv^\gP_i = \text{MLP}([ \vv^\gP_i, \text{Agg}([\text{MLP}(\vv^{\gP_1}_i), \cdots, \text{MLP}(\vv^{\gP_K}_i)]))$, where $\vv^{\gP_k}_i$ is the atom condition feature that corresponds to a specific row of $\mP_k$ if this atom belongs to the $k$-th subpocket and is set to be $\vzero$ otherwise.   
For the SE(3)-equivariant feature (i.e., the coordinate in 3D space) of each atom, we do not introduce any conditions.
Nevertheless, the geometric information is embedded in the SE(3)-invariant features thanks to the particular design of EGNN. After the input feature is augmented by the condition, the rest of the diffusion-based decoder mainly follows DecompDiff \citep{pmlr-v202-guan23a}, including the decomposed prior distribution, model architecture, training loss, etc.

% Our method also shares similarities with the variational autoencoder (VAE) \citep{kingma2013auto}, because it has the encoder-decoder architecture. However, there are three major differences from VAE. First, the decoder in our method is based on the more expressive diffusion model. The strong expressivity of diffusion models when serving as a decoder has been demonstrated in diffusion-based representation learning \citet{abstreiter2021diffusion,wang2023infodiffusion}. Second, 

\begin{wrapfigure}{R}{0.53\textwidth}
\begin{minipage}{0.53\textwidth}
% \vspace{-0.3in}
\begin{algorithm}[H]
% \caption{Optimization Process based on \method}
\caption{Optimization Process}
\label{alg:optimization}
\begin{algorithmic}
% \begin{algorithmic}[1]
\Require A specific protein pocket $\gP$ with detected subpockets $\{\gP_k\}_{k=1,\ldots,K}$, a reference ligand $\gM$, pre-trained decomposed and controllable diffusion model denoted as $\text{Enc}(\cdot)$ and $\text{DiffDec}(\cdot)$ 
\Ensure $\text{OAL}(\gP_k)$
% \State
\State \textcolor{teal}{\# Intialize all ordered arm lists for arms.}
\State $\{\gA_k\} \gets \text{Decompose}(\gM,\gP,\{\gP_k\})$ 
\For{$k \gets 1$ to $K$}
    \State \textcolor{teal}{\# $s_k$ is the evaluated score.}
    \State $(s_k, \gA_k) \gets \text{DockEval}(\gA_k, \gP_k)$
    \State $\text{OAL}(\gP_k)$ $\gets \langle(s_k, \gA_k)\rangle$
\EndFor
\State \textcolor{teal}{\# Start iterative optimization.}
\For{$ i \gets 1$ to $N$}
    \State \textcolor{teal}{\# The inner loop is just for better illustration} \State \textcolor{teal}{\# and is paralled as a batch in practice.}
    \For{$ j \gets 1$ to $B$}
        \State Sample $\{\gA_k\}$ from $\text{OAL}(\gP_k)$
        \State $\gM \gets \text{DiffDec}(\gP, \{\text{Enc}(\gA_k, \gP_k)\})$
        \State $\{\gA_k\} \gets \text{Decompose}(\gM, \gP, \{\gP_k\})$
        \For{$ k \gets 1$ to $K$}
        \State $(s_k, \gA_k) \gets \text{DockEval}(\gA_k, \gP_k)$
        \State Append $(s_k, \gA_k)$ to $\text{OAL}(\gP_k)$
        \State Sort $\text{OAL}(\gP_k)$ by $s_k$
        \State Keep top-$M$ elements in $\text{OAL}(\gP_k)$
        \EndFor
    \EndFor
\EndFor
\end{algorithmic}
\end{algorithm}
% \vspace{+0.6in}
\end{minipage}
\end{wrapfigure}

To shed lights on the insights behind the dedicated model design, we will discuss more about our special considerations as follows. The encoded conditions follow the principle of decomposition. It is notable that different reference arms that locally interact with different subpockets are encoded and play roles in parallel, which allows us to control the generation more flexibly. For example, we can control the different arms of the generated ligand molecules separately. Another concern is about diversity. We do not explicitly introduce any regularization like VAE \citep{kingma2013auto} over the representation space of conditions in consideration of the unsmooth nature of chemical space. Nevertheless, the source of randomness is two-fold: the sampling procedure of the diffusion model and the degree of freedom of scaffold.   
% Thus the space of the representation may not be smooth. Our consideration is that the chemical space itself is far from smooth. \xiangxin{mention activity cliffs or not? It may be conflict with some ideas in the next section.} \xiangxin{think up proper examples for no smoothness in chemical space} For example, activity cliffs \citep{cruz2014activity} are a well-known concept in drug discovery. Pairs of compounds with high structural similarity may be highly different in activity or other properties. Even without a probabilistic latent space, 
Notably, the scaffold and arms will impact each other in the generative process and the randomness of scaffolds will also flow into arms. Expectedly, the arm of the generated ligand molecule will be similar to its corresponding reference arm but not exactly the same, which is the workhorse of our framework. This characteristic reflects both the abilities of exploration and exploitation, which is critical to the optimization process that will be introduced in the following. 
% However, lack of smoothness in the representation space of conditions, we cannot leverage gradients or any other higher-order information for optimization. Thus we propose a zero-order optimization particularly designed for our model, which will be introduced in \cref{subsec:optimization_procedure}.
\clearpage
\subsection{Optimization in the Decomposed Drug Space}
% \subsection{\mbox{Optimization in the Decomposed Drug Space}}
\label{subsec:optimization_procedure}

Thanks to the controllability and decomposition of the model introduced above, we can optimize the ligand molecules in the decomposed drug space. We will introduce the optimization process of \method as follows.

Due to the characteristics of decomposition, different arms that locally  interact with different subpockets can be evaluated seperately. Thus we can define a score for each arm, which can be a single objective or a value that scalarizes multiple objectives by weighted sum.
The following optimization process is oriented towards a given protein pocket. For each subpocket of the given protein pocket, we build an ordered list with a certain max size to store potential arms sorted by their scores. We can initialize the ordered arm lists (OAL) by decomposing reference ligands or ligand molecules generated by generative models.
In each iteration of optimization, we use the controllable and decomposed diffusion models to generate a batch of ligand molecules conditioned on reference arms sampled from the ordered arm lists and further decompose them to get new arm candidates. The new arms are first refined by re-docking and then evaluated by oracles. Here we introduce an optional re-docking procedure to assign each arm with a higher-quality pose so that better interaction information can be injected into arm conditions. Then the new arms are inserted into the corresponding ordered lists and the arms with bad scores will be remove to keep predefined max size. As the process goes on, more and more arms that can well interact with the subpockets will be discovered. 

The optimization process based \method is summarized as \cref{alg:optimization}. This optimization process shares similar ideas with well-recognized evolutionary algorithms. Arms in each ordered arm list evolve towards desired properties. The controllable generation can be viewed as one kind of mutation. But, differently, generative mutation is more efficient than the widely-used mutation defined by rule-based perturbation.

\section{Experiments}

\subsection{Experimental Setup}
\label{subsec:experimental_setup}
\paragraph{Dataset} 
We utilized the CrossDocked2020 dataset \citep{francoeur2020three} to train and evaluate our model. Additionally, we adopted the same filtering and splitting strategies as the previous work \citep{luo20213d, peng2022pocket2mol,guan3d}. The strategy focuses on retaining high-quality complexes (RMSD $<1$\AA) and diverse proteins (sequence identity $<$ $30\%$), leading to $100,000$ protein-binding complexes for training and  $100$ novel protein for testing. 

\noindent\textbf{Implementation Details} 
For iterative optimization, we select arms with desired properties as conditions. \revise{We initialize the list of arms with 20 molecules generated by DecompDiff in our experiment.} To better align with the practical requirements of pharmaceutical practice, where the goal is to generate molecules with high drug likeness, synthesis feasibility, binding affinity, and other pertinent properties, we introduced a multi-objective optimization score to effectively balance these different objectives. In each iteration, we evaluate QED, SA, Vina Min (which will be introduced in detail later) of decomposed arms from generated molecules, then calculate $\text{Z-score}(x_{i}) = (x_{i} - \text{mean}(X)) / \text{std}(X), \ x_{i}\in X$ (also known as the standard score \citep{zill2020advanced}) of each property, where $X$ denotes the set of evaluated values of a specific property. The Z-scores of each property are aggregated with same weights as criteria for selecting the top-$M$ arms as next iteration conditions. In our experiment, we conduct 30 rounds of optimization and sample 20 molecules in each round. For top-$M$ arm selection, we set $M$=3. For each protein pocket, if the average properties of sampled ligand molecules are no longer improved, the optimization process will be early stopped.

\noindent\textbf{Baselines} There are two types of baselines from the perspective of generation and optimization, respectively. \textbf{Generation Perspective}: We compare our model with various representative generative baselines: \textbf{liGAN} \citep{ragoza2022generating} is a 3D CNN-based conditional VAE model which generate ligand molecules in atomic density grids. \textbf{AR} \citep{luo20213d}, \textbf{Pocket2Mol} \citep{peng2022pocket2mol} and \textbf{GraphBP} \citep{liu2022generating} are GNN-based methods that generate 3D molecules atom by atom in an autoregressive manner. \textbf{TargetDiff} \citep{guan3d} is a diffusion-based method which generates atom coordinates and atom types in a non-autoregressive way, and the prior distribution is a standard Gaussian and bonds are generated with OpenBabel \citep{o2011open}. \textbf{DecompDiff} \citep{pmlr-v202-guan23a} is a diffusion-based method with decomposed priors and validity guidance which generates atoms and bonds of 3D ligand molecules in an end-to-end manner. Decompdiff has three optional decomposed priors: reference priors, pocket priors, and optimal priors. Our method also follows this setting. \textbf{Optimization Perspective}: We choose the most related work, \textbf{RGA} \citep{fu2022reinforced}, which is a reinforced genetic algorithm with policy networks and also focus on structure-based molecular optimization, as the baseline. Besides, we also introduced the controllability into TargetDiff by whole molecule conditions rather than arm conditions. We name this baseline as \textbf{TargetDiff w/ Opt.}.

\begin{table*}[t!]
    % \vspace{-0.05in}
    \setlength{\extrarowheight}{1.7pt}
    \centering
    \caption{Summary of different properties of reference molecules and molecules generated by our model and other generation (Gen.) and optimization (Opt.) baselines. ($\uparrow$) / ($\downarrow$) denotes a larger / smaller number is better. Top 2 results are highlighted with \textbf{bold text} and \underline{underlined text}, respectively. 
    }
    % \vspace{-0.07in}
    \begin{adjustbox}{width=1\textwidth}
    \renewcommand{\arraystretch}{1.2}
    \begin{tabular}{l|c|cc|cc|cc|cc|cc|cc|cc|c}
    \toprule
    % \diagbox{Model}{Metric} 
    \multicolumn{2}{c|}{\multirow{2}{*}{Methods}} & \multicolumn{2}{c|}{Vina Score ($\downarrow$)} & \multicolumn{2}{c|}{Vina Min ($\downarrow$)} & \multicolumn{2}{c|}{Vina Dock ($\downarrow$)} & \multicolumn{2}{c|}{High Affinity ($\uparrow$)} & \multicolumn{2}{c|}{QED ($\uparrow$)}   & \multicolumn{2}{c|}{SA ($\uparrow$)} & \multicolumn{2}{c|}{Diversity ($\uparrow$)} & \multicolumn{1}{c}{Success} \\
    \multicolumn{2}{c|}{} & Avg. & Med. & Avg. & Med. & Avg. & Med. & Avg. & Med. & Avg. & Med. & Avg. & Med. & Avg. & Med. & Rate ($\uparrow$)\\
    
    \toprule
    
    \multicolumn{2}{c|}{Reference}   & -6.36 & -6.46 & -6.71 & -6.49 & -7.45 & -7.26 & -  & - & 0.48 & 0.47 & 0.73 & 0.74 & - & - & 25.0\%  \\
   
    \midrule
    
    \multirow{6}{*}{Gen.} & LiGAN       & - & - & - & - & -6.33 & -6.20 & 21.1\% & 11.1\% & 0.39 & 0.39 & 0.59 & 0.57 & 0.66 & 0.67 & 3.9\% \\
    
    & GraphBP     & - & - & - & - & -4.80 & -4.70 & 14.2\% & 6.7\% & 0.43 & 0.45 & 0.49 & 0.48 & \textbf{0.79} & \textbf{0.78} & 0.1\%  \\
    
    & AR          & -5.75 & -5.64 & -6.18 & -5.88 & -6.75 & -6.62 & 37.9\% & 31.0\% & 0.51 & 0.50 & 0.63 & 0.63 & 0.70 & 0.70 & 7.1\% \\
    
    & Pocket2Mol  & -5.14 & -4.70 & -6.42 & -5.82 & -7.15 & -6.79 & 48.4\% & 51.0\% & 0.56 & \underline{0.57} & \textbf{0.74} & \textbf{0.75} & 0.69 & 0.71 & 24.4\% \\
    
    & TargetDiff  & -5.47 & -6.30 & -6.64 & -6.83 & -7.80 & -7.91 & 58.1\% & 59.1\% & 0.48 & 0.48 & 0.58 & 0.58 & \underline{0.72} & \underline{0.71} & 10.5\% \\
    
    & DecompDiff & -5.67 & -6.04 & -7.04 & -7.09 & \underline{-8.39} & \underline{-8.43} & 64.4\% & 71.0\% & 0.45 & 0.43 & 0.61 & 0.60 & 0.68 & 0.68 & 24.5\% \\
    
    \midrule
    
    \multirow{1}{*}{Opt.} & RGA  & - & - & - & - &  -8.01 & -8.17 & 64.4\% & 89.3\% & \underline{0.57} & \underline{0.57} & \underline{0.71} & \underline{0.73} & 0.41 & 0.41 & \underline{46.2\%} \\
    \midrule

    \multirow{0.5}{*}{\thead{Gen.\\ + \\ Opt.}}

    & TargetDiff w/ Opt. & \textbf{-7.87} & \textbf{-7.48} & \textbf{-7.82} & \underline{-7.48} & -8.30 & -8.15 & \underline{71.5\%} & \textbf{95.9\%} & \textbf{0.66} & \textbf{0.68} & 0.68 & 0.67 & 0.31 & 0.30 & 25.8\% \\
    
    & \method & \underline{-5.87}  & \underline{-6.81} & \underline{-7.35} & \textbf{-7.72} & \textbf{-8.98} & \textbf{-9.01} & \textbf{73.5\%} & \underline{93.3\%} & 0.48 & 0.45 & 0.65 & 0.65 & 0.60 & 0.61 & \textbf{52.5\%} \\

    \bottomrule
    \end{tabular}
    \renewcommand{\arraystretch}{1}
    \end{adjustbox}\label{tab:main_tab}
    % \vspace{-0.1in}
\end{table*}
\paragraph{Metircs} 
We evaluate the generated molecules from \textbf{target binding affinity and molecular properties}. 
% The premise for SBDD is that the model could generate accurate ligand conformations to demonstrate an understanding of molecular validity and stability.
% we first plot the carbon-carbon bond distance distribution and all-atom pairwise distance distribution of the generated molecules.
% in Figure \ref{fig:atom_jsd}. 
We employ AutoDock Vina \citep{eberhardt2021autodock} to estimate the target binding affinity, following the same setup as \citet{luo20213d, ragoza2022generating}. 
We collect all generated molecules across 100 test proteins and report the mean and median of affinity-related metrics~(\textit{Vina Score}, \textit{Vina Min}, \textit{Vina Dock}, and \textit{High Affinity}) and property-related metrics~(drug-likeness \textit{QED} \citep{bickerton2012quantifying}, synthesizability \textit{SA} \citep{ertl2009estimation}, and \textit{diversity}). 
Vina Score directly estimates the binding affinity based on the generated 3D molecules, Vina Min conducts a local structure minimization before estimation, Vina Dock involves a re-docking process and reflects the best possible binding affinity, and High Affinity measures the percentage of how many generated molecules binds better than the reference molecule per test protein.
% Therefore, \textit{Vina Score} and \textit{Vina Min} are more direct metrics for evaluating the end-to-end 3D generation performance
Following \citet{yang2021knowledge, long2022zero,pmlr-v202-guan23a,mars3d}, we further report the percentage of molecules which pass certain criteria (QED $> 0.25$, SA $> 0.59$, Vina Dock $< -8.18$) as \textit{Success Rate} for comprehensive evaluation, considering that practical drug design also requires the generated molecules to be drug-like, synthesizable, and maintain high binding affinity simultaneously \citep{jin2020multi, xie2021mars}.
% The thresholds used for QED and SA are computed as the 10th percentile of molecules in the DrugCentral database \cite{ursu2016drugcentral}, which are all pharmaceutical or under clinical trials.
Additionally, we also evaluate the generated molecules from the perspective of molecular conformation. Please refer to \cref{appendix:molecular_conformation} for more results.

\subsection{Main Results}

We evaluate the effectiveness of our model in terms of binding affinity and molecular properties. 
As shown in \cref{tab:main_tab}, \method outperforms baselines by a large margin in affinity-related metrics and Success Rate, a comprehensive metric. 
Specifically, \method surpasses the strong baseline DecompDiff in all metrics, except diversity. 
% Specifically, \method surpasses the strong baseline DecompDiff in almost all metrics. 
All these gains clearly indicate that the optimization works as we expected and our method better aligns with the goal of drug discovery compared with the generative baselines.
% All these gains clearly indicate that the optimization works as expected and better aligns with the goal of drug discovery compared with the generative baselines.

Notably, RGA, which also focuses on structure-based molecular optimization, achieves promising Sucess Rate. And our method performs even better. More importantly, molecular optimization methods like RGA inevitably encounter the pitfalls of low diversity and inefficiency. Because these optimization methods usually start with a reference ligand and iteratively make small changes on it to gradually improve the property. They are easy to be trapped in local solutions. However, \method inherites advantages of both optimization and generation, achieving high Success Rate and considerable diversity at the same time. See \cref{appendix:additional_results} for additional results.
\definecolor{darkgreen}{RGB}{34,139,34}
\begin{table}
\parbox{0.60\linewidth}{
\centering
\fontsize{8}{9.5}\selectfont
\caption{Summary of results of single-objective optimization. The improvements of \method over the baseline are highlighted in \textcolor{darkgreen}{green}.}
    \renewcommand{\arraystretch}{1.3}
    \begin{tabular}{c|cc|cc}
    \toprule
    % \diagbox{Model}{Metric} 
    \multirow{2}{*}{Property}
    & \multicolumn{2}{c|}{DecompDiff} & \multicolumn{2}{c}{\method} \\
    & Avg. & Med. & Avg. & Med. \\ 
    \midrule
    QED ($\uparrow$) & 0.48 & 0.48 & 0.52 \textcolor{darkgreen}{(+8.3\%)} & 0.53 \textcolor{darkgreen}{(+10.4\%)} \\
    SA ($\uparrow$) & 0.67 & 0.66 & 0.74 \textcolor{darkgreen}{(+10.5\%)} & 0.74 \textcolor{darkgreen}{(+12.1\%)} \\
    Vina Min ($\downarrow$) & -6.01 & -5.98 & -6.72 \textcolor{darkgreen}{(+11.8\%)} & -6.72 \textcolor{darkgreen}{(+12.4\%)} \\
    \bottomrule
    \end{tabular}
    \label{tab:single_objective_table}
    
}
\hfill 
\parbox{.35\linewidth}{
\centering 
    \caption{Comparison of optimization strategies using molecule-level and arm-level conditions.}
    \fontsize{8}{9.5}\selectfont
    %\vspace{-0.07in}
    % \begin{adjustbox}{width=1\textwidth}
    \renewcommand{\arraystretch}{1.3}
    \begin{tabular}{c|cc}
    \toprule
    % \diagbox{Model}{Metric} 
    \multirow{2}{*}{Method} 
    & \multicolumn{2}{c}{Vina Min ($\downarrow$)} \\
    & Avg. & Med. \\ 
    \midrule
    DecompDiff & -6.01 & -5.98 \\
    Molecule-level Opt. & -6.62 & -6.66 \\
    Arm-level Opt. & -6.72 & -6.72 \\
    \bottomrule
    \end{tabular}\label{tab:mol_wise_table}
}
\end{table}

\subsection{Ablation Studies}

\paragraph{Single-Objective Optimization} To further validate the effectiveness of \method, we test our method on the setting of single-objective optimization with reference priors. Specifically, we use QED, SA, and Vina Min as the objective and analyze the results of three experiments, respectively.   
As shown in \cref{tab:single_objective_table}, each experiment effectively improves the corresponding property.

%\paragraph{Diversity} \xiangxin{need yuwei's help to emphasize the importance of diversity in drug discovery} Here, we want to show the trade-off between Success Rate and diversity of RGA, TargetDiff w/ Opt., and \method for each target protein pocket. As shown in \cref{fig:diversity_scatter}, \method shows general superiority to the other two baselines in most cases considering both Success Rate and diversity. 

\noindent\textbf{Benefits of Decomposed Optimization} The decomposed drug space allows for optimizing each arm of a  ligand molecule separately. In our method, arms of each subpockets are evaluated and selected independently (namely, arm-level optimization). We also tried evaluating the property of the \underline{whole} ligand molecules, choosing those with desired property, and decomposing them to serve as the arm conditions in the next optimization iteration (namely, molecule-level optimization). We compare these two optimization under the setting of reference prior. As shown in \cref{tab:mol_wise_table}, arm-wise 
optimization performs better than molecule-wise optimization, which demonstrates benefits brought by decomposition in terms of optimization efficiency. 
\subsection{Controllability}

Various molecular optimization scenarios, including R-group design and scaffold hopping, play a crucial role in real-world drug discovery. They enhance binding affinity, potency, and other relevant molecular properties with greater precision. Our controllable and decomposed diffusion model seamlessly integrates with these scenarios by incorporating expert knowledge through decomposed arm conditions, better aligned with the demands and objectives of the pharmaceutical industry.
\begin{table*}[ht]
    % \vspace{-0.05in}
    \small
    \centering
    \caption{Scaffold hopping results 
    % .generated 
    of Decompdiff and \method on CrossDocked2020 test set.}
    % \vspace{-0.07in}
    \begin{adjustbox}{width=0.8\textwidth}
    \renewcommand{\arraystretch}{1.2}
    \begin{tabular}{c|c|c|c|c|c}
    \toprule
    % \diagbox{Model}{Metric} 
    \multicolumn{1}{c|}{\multirow{2}{*}{Methods}}
    & \multicolumn{1}{c|}{\multirow{2}{*}{\thead{Valid\\($\uparrow$)}}} 
    & \multicolumn{1}{c|}{\multirow{2}{*}{\thead{Unique\\($\uparrow$)}}}
    & \multicolumn{1}{c|}{\multirow{2}{*}{\thead{Novel\\($\uparrow$)}}}
    & \multicolumn{1}{c|}{\multirow{2}{*}{\thead{Complete\\~~Rate ($\uparrow$)}}}
    & \multicolumn{1}{c}{\multirow{2}{*}{\thead{Scaffold\\Similarity ($\downarrow$)}}} \\ 
    & & & & & \\
    \midrule
    DecompDiff+Inpainting & 0.95 & 0.48 & 0.85 & 89.2\% & 0.40 \\
    \method+Inpainting & 0.96 & 0.46 & 0.88 & 93.0\% & 0.35 \\
    \bottomrule
    \end{tabular}\label{tab:scaffold_hopping}
    \renewcommand{\arraystretch}{1}
    \end{adjustbox}
    % \vspace{-0.1in}
\end{table*}

\paragraph{R-group Design}
R-group optimization is a widely used technique to optimize molecules' substituents for improving biological activity. Our model is well-suited for this task by employing finer-level arm conditions to guide the optimization. To achieve the optimization goal, we decompose the compound into a scaffold and multiple arms. Subsequently, we choose a specific arm for optimization to enhance its binding affinity. The optimization process involves conditional inpainting \citep{lugmayr2022repaint} inspired by \citet{schneuing2022structure}.

We diffuse the remaining part of the compound while predicting the selected arm, which is conditioned on the reference arm at each step. \revise{We initialize the list of arms by the arms defined as r-group to optimize.} From the molecules generated through this process, we can select substituents with higher \revise{Vina Min Score} to serve as the condition for the next iteration. Results of R-group optimization on protein 3DAF and 4F1M are presented in \cref{fig:R-group}. \revise{More results on protein 4G3D can be found in \cref{appendix:controllability}.} After 30 rounds of optimization, our generated molecules achieve a vina minimize score more than 1 kcal/mol better than the reference. Moreover, we compare \method with Decompdiff for R-group optimization on Vina minimize, Tanimoto similarity, and Complete rate with detailed result in \cref{appendix:controllability}. Tanimoto similarity is calculated using rdkit \texttt{GetMorganFingerprint}
 and \texttt{TanimotoSimilarity} functions. Complete rate measures the proportion of \revise{completed molecules in} generated result. As \cref{tab:r-group_eval} shows, the decomposed conditional arms bring greater controllability over the shape, positioning, and properties of the generated substituents compared to diffusion inpainting.

Fragment growing is a another technique for R-group design. Unlike R-group optimization, fragment growing aims to design new substituents instead of optimization existing ones. 
%is a commonly employed technique to enhance the drug candidacy of molecules displaying favorable properties. 
By designing novel arms priors and predicting the number of atoms through chemistry software or expert guidance, \method can naturally facilitate incremental growth and optimization of newly generated arms, leading to improved biological activity. A case study on the application of fragment growing can be found in \cref{appendix:controllability}.

\paragraph{Scaffold Hopping}
\begin{wrapfigure}{r}{0.62\textwidth}
    %\begin{tabular}{@{}c@{}}
    %\centering
    % \vspace{-20pt}
    \includegraphics[width=0.62\textwidth]{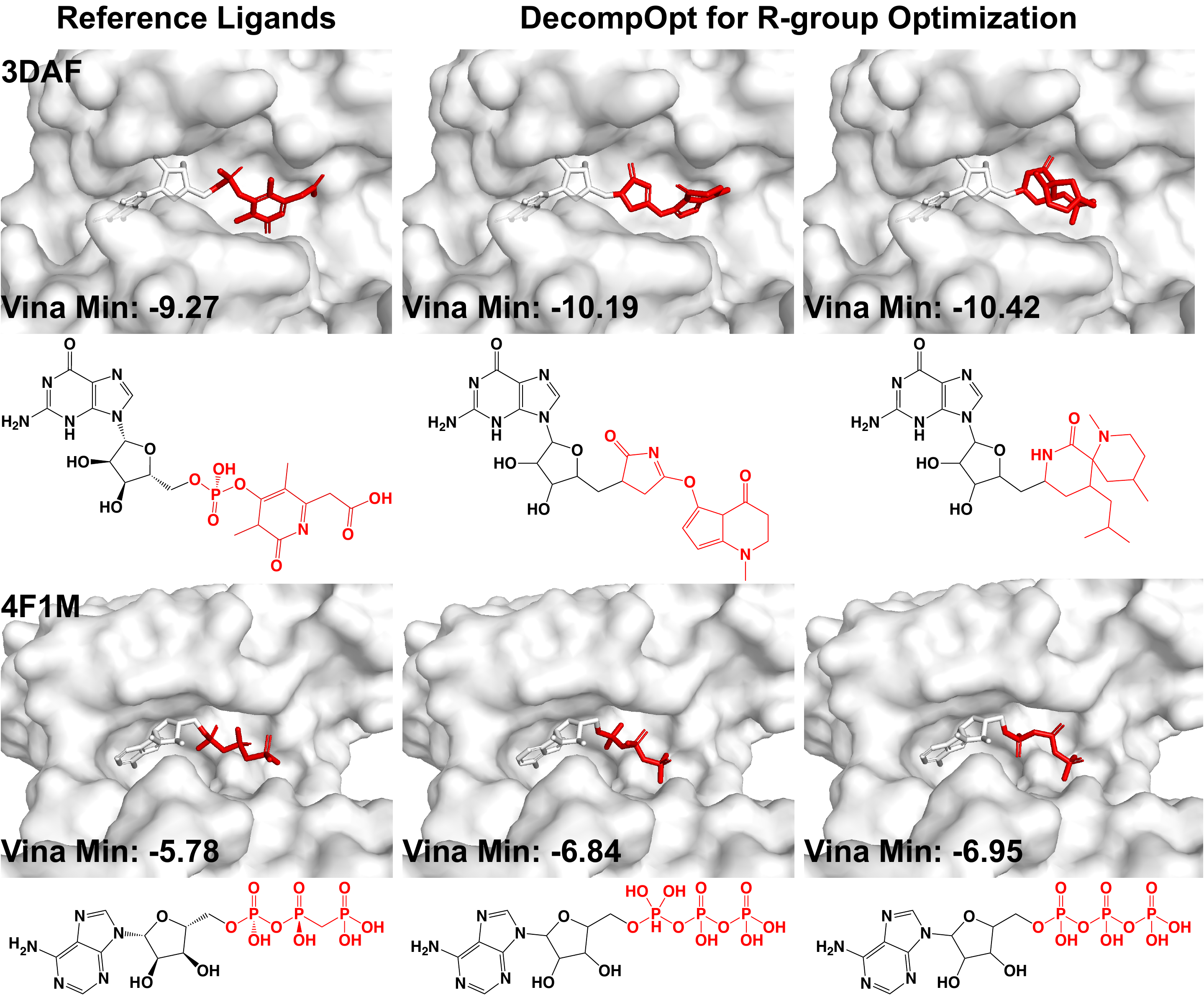}
    \captionof{figure}{\revise{Visualization of reference binding molecules (left column), molecules generated by \method (middle and right column) with 30 rounds of optimization on protein 3DAF (top row) and 4F1M (bottom row). Optimized R-group are highlighted in red.}}
    \label{fig:R-group}
    \vspace{+0.20in}
    \smallskip\par
    \includegraphics[width=0.62\textwidth]{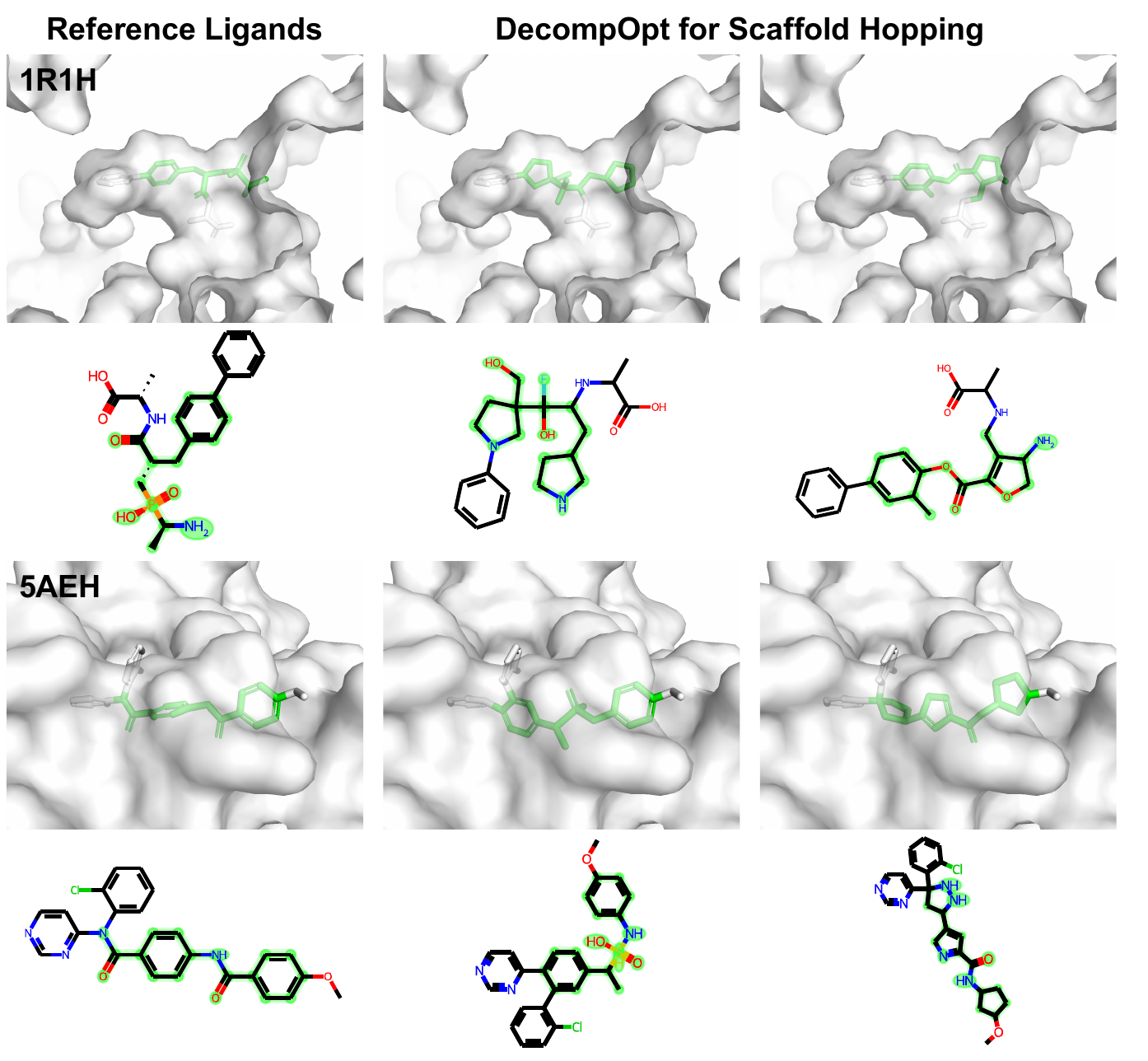}
    \captionof{figure}{Examples of scaffold hopping accomplished by \method. For each row, the left image shows the reference ligand, the middle and right images are two examples generated by \method. Reference and generated scaffolds are highlighted in green.}
    \label{fig:scaffold_hopping}
    \vspace{-20pt}
    % \textcolor{blue}{\rule{3cm}{3cm}} \\% Dummy image replacement
    %\textcolor{green}{\rule{3cm}{3cm}} \\% Dummy image replacement
    
    %\end{tabular}
\end{wrapfigure}
In contrast to R-group design, scaffold hopping involves the generation of scaffolds to connect pivotal groups, which play a critical role in interaction with proteins. We apply our decomposed conditional diffusion model to scaffold hopping by inpainting scaffold in fixed arms and incorporating pivotal arms as structural conditions. We generate 20 molecules for each target on our test set. Following \citet{huang20223dlinker}, we measure the \textit{Validity}, \textit{Uniqueness}, and \textit{Novelty} of generated molecules. Additionally, we compute the \textit{Complete Rate}, which measures the proportion of successfully constructed molecules with all atoms connected. To better understand the influence of conditional arms on scaffold generation, we estimate \textit{Scaffold Similarity} between generated scaffold and reference scaffold following \citet{polykovskiy2020molecular}. A detailed description of scaffold hopping evaluation metrics can be found in \cref{appendix:controllability}.
The supplementary information provided indicates the inclusion of arm conditions can influence scaffold generation through message passing, leading to a more consistent scaffold when compared to diffusion inpainting without arm conditions. As \cref{tab:scaffold_hopping} shows, our model achieves higher validity and complete rate. Higher novelty and lower scaffold similarity indicate that our model is better at maintaining diversity and exploring novel molecules while controlling pivotal groups. 
In \cref{fig:scaffold_hopping}, we show results of scaffold hopping on 1R1H and 5AEH. 

More visualization results can be found in \cref{appendix:controllability}.

\section{Conclusions}

In this work, we proposed a controllable and decomposed diffusion model for structure-based molecular optimization and opened a new paradigm that combines generation and optimization for structure-based drug design. Our method shows promising performance on both \textit{de novo} design and controllable generation, indicating its great potential in drug discovery. We would like to point out that in our current controllable diffusion models, we did not explore the best way for multi-objective optimization. We plan to  investigate it in our future work. 
\pagebreak
\bibliography{main}
\bibliographystyle{iclr2024_conference}

\clearpage
\appendix

\revise{\section{Implementation Details}}

\revise{In this section, we will provide more implementation details of our methods. Though some contents, such as the SE(3)-equivariant layer and the loss function, can be referred to \citet{pmlr-v202-guan23a}, we still include them here to make our paper more self-containing.}

\revise{\subsection{Featurization}}
\revise{
We follow the decomposition algorithm proposed by \citet{pmlr-v202-guan23a} to decompose ligand molecules into arms and a scaffold. We define the part of proteins that lies within 10$\text{\AA}$ of any atom of an arm as its corresponding subpocket. }

\revise{
Following DecompDiff \citep{pmlr-v202-guan23a}, we represent each protein atom with the following features: one-hot element indicator (H, C, N, O, S, Se), one-hot amino acid type indicator (20 dimension), one-dim flag indicating whether the atom is a backbone atom, and one-hot arm/scaffold region indicator. 
If the distance between the protein atom and any arm center is within 10$\text{\AA}$, the protein atom will be labeled as belonging to an arm region and otherwise a scaffold region. 
The ligand atom is represented with following features: one-hot element indicator (C, N, O, F, P, S, Cl) and one-hot arm/scaffold indicator. Different from DecompDiff, the atom features are enhanced by concatenating an SE(3)-invariant feature of arms and their corresponding subpockets encoded by the condition encoder after the orignal features.
}

\revise{
Two graphs are constructed for message passing in the protein-ligand complex: a k-nearest neighbors graph $\mathcal{G}_K$ upon ligand atoms and protein atoms (we choose $k=32$ in all experiments) and a fully-connected graph $\gG_L$ upon ligand atoms. 
The edge features are the outer products of distance embedding and edge type. 
The distance embedding is obtained by expanding distance with radial basis functions (RBF) located at 20 centers between 0$\text{\AA}$ and 10$\text{\AA}$. The edge type is a 4-dim one-hot vector indicating the edge is between ligand atoms, protein atoms, ligand-protein atoms or protein-ligand atoms. In the ligand graph, the ligand bond is represented with a one-hot bond type vector (non-bond, single, double, triple, aromatic), an additional feature indicating whether or not two ligand atoms are from the same arm/scaffold.
}

\revise{
\subsection{Model Details}
}

\revise{
The controllable and decomposed diffusion model consist of two parts: a condition encoder and a diffusion-based decoder. The building block is an SE(3)-equivariant layer that is composed of three layers: atom update layer, bond update layer, and position update layer. 
}

\revise{
We denote the protein pocket as $\gP=\{(\rvx^\gP_i,\rv^\gP_i)\}_{i\in\{1,\dots,N_\gP\}}$ and the ligand molecule as $\mathcal{M}=\{(\vx_i, \vv_i, \vb_{ij})\}_{i,j\in\{1,\dots,N_\mathcal{M}\}}$, where $\vx$ is the atom position, $\vv$ is the atom type, and $\vb_{ij}$ is the chemical bond type between the atom $i$ and the atom $j$. For brevity, we omit the superscript $\gP$ or $\gM$ in the following. We use $\rvh_i$ to denote the the SE(3)-invariant hidden state of $i$-th atom, $\rvx_i$ to denote the $i$-th atom's coordinate, which is SE(3)-equivariant, and $\rve_{ij}$ to denote the hidden state of the edge between the $i$-th atom and the $j$-th atom. They can be obtained as we described in the previous subsection. And we use $t$ to denote the time embedding as that in \citet{ho2020denoising}.
}

\revise{
\paragraph{Atom Update Layer} We denote the atom update layer as $\phi_a\coloneqq \{\phi_{a1},\phi_{a2},\phi_{a3},\phi_{a4}\}$.
}

\revise{
We first use the atom update layer $\phi_{a1}$ to model protein-ligand interation as follows:
\begin{equation}
    \Delta \rvh_{K,i} \leftarrow \sum_{j\in \mathcal{N}_K(i)}\phi_{a1}(\rvh_i,\rvh_j,||\rvx_i-\rvx_j||, \rve_{ij}, t), 
\end{equation}
where $\gN_K(i)$ is the set of neighbors of the $i$-th atom in the protein-ligand complex graph $\gG_K$.
}

\revise{
We then further use the atom update layer $\phi_{a2}$ and $\phi_{a3}$ to model the interaction inside the ligand as follows:
\begin{align}
    & \rvm_{ij} \leftarrow \phi_{a2}(||\rvx_i-\rvx_j||, \rve_{ij}), \\
    & \Delta \rvh_{L,i} \leftarrow \sum_{j\in\gN_L(i)} \phi_{a3}(\rvh_i, \rvh_j, \rvm_{ji},t),
\end{align}
where $\gN_L(i)$ represents the set of neighbors of the $i$-th atom in the ligand graph $\gG_L$.
}
\revise{
Finally, we update the hidden state of atoms by the atom update layer $\phi_{a4}$ as follows:
\begin{equation}
    \rvh_i\leftarrow \rvh_i + \phi_{a4}(\Delta\rvh_{K,i} + \Delta\rvh_{L,i}).
\vspace{-6mm}
\end{equation}}

\revise{\paragraph{Bond Update Layer} We update the hidden states of the edges by the bond update layer $\phi_b$ as follows:
\begin{equation}
    \rve_{ij}\leftarrow\sum_{k\in\gN_L(i)\backslash \{j\}}\phi_{b}(\rvh_i,\rvh_j,\rvh_k,\rvm_{kj},\rvm_{ji},t).
    \vspace{-6mm}
\end{equation}}

\revise{
\paragraph{Position Update Layer} The atom positions are updated by the position update layer $\phi_{p}\coloneqq \{\phi_{p1},\phi_{p2}\}$ as follows:
\begin{align}
    &\Delta \rvx_{K,i} \leftarrow \sum_{j\in\gN_K(i)} (\rvx_j-\rvx_i)\phi_{p1}(\rvh_i,\rvh_j,||\rvx_i-\rvx_j||,t), \\
    &\Delta \rvx_{L,i} \leftarrow \sum_{j\in\gN_L(i)} (\rvx_j-\rvx_i)\phi_{p2}(\rvh_i,\rvh_j,||\rvx_i-\rvx_j||,\rvm_{ji},t), \\
    & \rvx_i \leftarrow \rvx_i + (\Delta \rvx_{K,i} + \Delta \rvx_{L,i})\cdot \mathds{1}_{\text{mol}},
\end{align}
where $\mathds{1}_{\text{mol}}$ is the indicator of ligand atoms since we assume the protein atoms are fixed as the context. 
}

\revise{
In practice, the condition encoder consists of two SE(3)-equivariant layers and the diffusion-based decoder consists of six SE(3)-equivaraint layers. In each SE(3)-equivariant layer, following \citet{pmlr-v202-guan23a}, we apply graph attention to aggregate the message of each node/edge. The key/value/query embedding is obtained through a 2-layer MLP with LayerNorm and ReLU activation. Stacking these three layers as a block, our model consists of 6 blocks with \texttt{hidden\_dim=128} and \texttt{n\_heads=16}. Additionally, the diffusion-based decoder also have two prediction heads (which are simply 2-layer MLPs and following Softmax function) that maps the learned hidden states of atoms and edges to the predicted atom type and bond type.
}
\revise{\subsection{Training Details}}

\begin{figure*}[h]
\centering
% \raggedleft
% \flushleft
\includegraphics[width=0.86\textwidth]{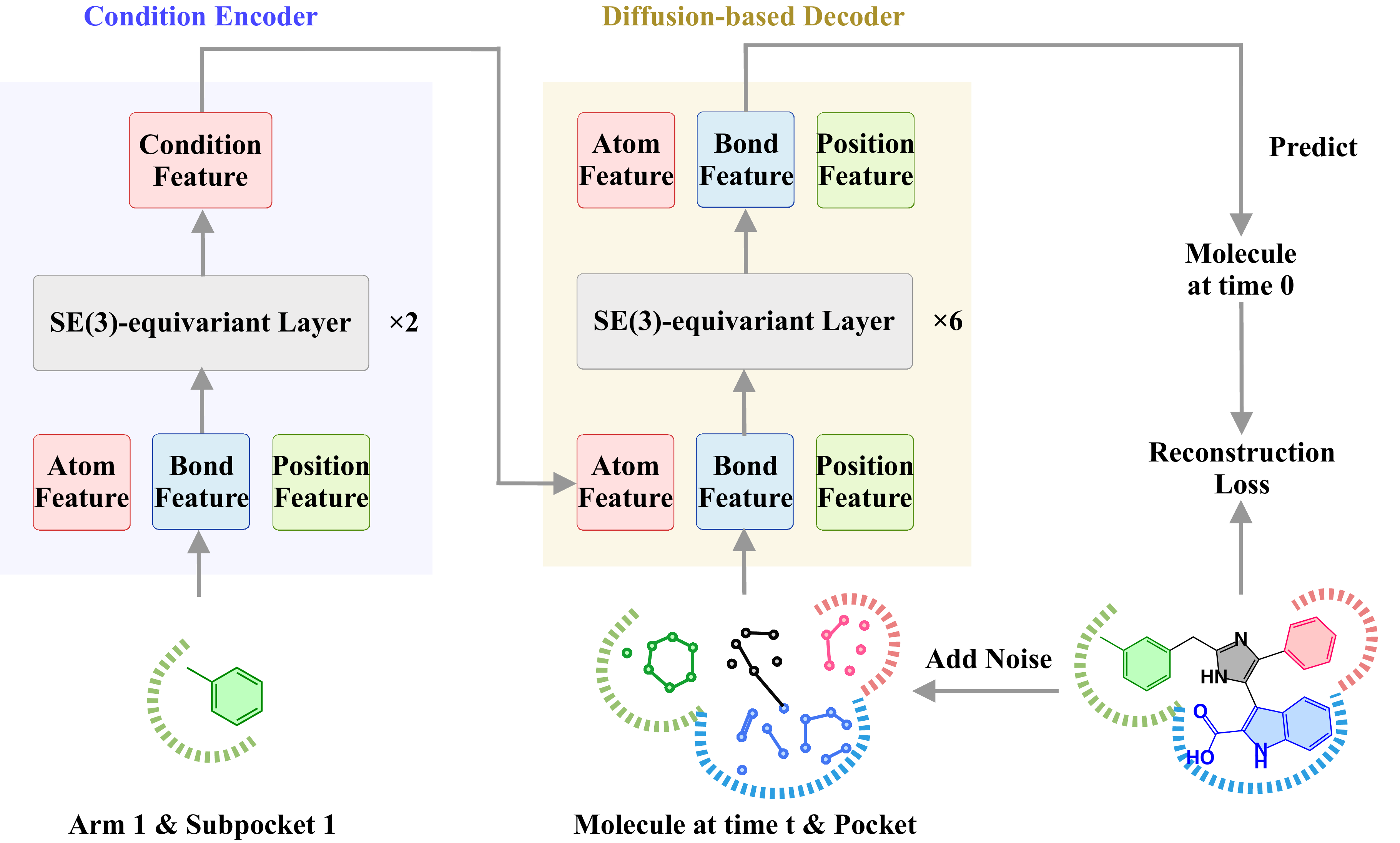}
  \caption{\revise{Illustration of training. For this case, there are actually three pairs of arms and subpockets input to the condition encoders separately. For brevity, we only plot one as an example.}}{\label{fig:training_figure_appendix}}
\end{figure*}

\revise{
Given a pair of protein and ligand molecule, we first decompose the molecule to get the arms. 
We add noise to the ligand molecules in the training set to get the perturbed molecules as the forward process of diffusion models (\eqref{eq:forward_process}). The forward process is a Markov chain with fixed variance schedule $\{\beta_t\}_{t=1,\dots,T}$ \citep{ho2020denoising}. We denote $\alpha_t=1-\beta_t$ and $\Bar{\alpha}_t=\prod_{s=1}^t\alpha_s$. 
More specifically, the noises at time $t$ are injected as follows:
\begin{align}
    & q(\rvx_t|\rvx_0) = \gN(\rvx_t; \sqrt{\Bar{\alpha}_t}\rvx_0,(1-\Bar{\alpha}_t)\rmI), \\
    & q(\rvv_t|\rvv_0) = \gC(\rvv_t| \Bar{\alpha}_t\rvv_0 + (1-\Bar{\alpha}_t)/K_a), \\
    & q(\rvb_t|\rvb_0) = \gC(\rvv_t| \Bar{\alpha}_t\rvb_0 + (1-\Bar{\alpha}_t)/K_b),
\end{align}
where $K_a$ and $K_b$ are the number of atom classes and bond classes respectively.
}

\revise{
Then the arms and subpockets are input to the condition encoder. The output of condition encoder and the perturbed ligand are further input to the diffusion-based decoder. Then the reconstruction loss $L_{t}$ at time $t$ is defined as follows:
\begin{align}
    & L_{t}^{(v)} = \sum_{k=1}^{K_a} \bm{c}(\rvv_t, \rvv_0)_k \log \frac{\bm{c}(\rvv_t, \rvv_0)_k}{\bm{c}(\rvv_t, \hat\rvv_0)_k}, \\
    & L_{t}^{(b)} = \sum_{k=1}^{K_b} \bm{c}(\rvb_t, \rvb_0)_k \log \frac{\bm{c}(\rvb_t, \rvb_0)_k}{\bm{c}(\rvb_t, \hat\rvb_0)_k}, \\ 
    & L_{t}^{(x)} = ||\rvx_0-\hat{\rvx}_0||^2, \\
    & L_{t} = L_{t}^{(x)} + \gamma_v L_{t}^{(v)} + \gamma_b L_{t}^{(b)},
\end{align}
where $(\rvx_t,\rvv_t,\rvb_t)$, $(\rvx_0,\rvv_t,\rvb_t)$, and $(\hat{\rvx}_0,\hat{\rvv}_0,\hat{\rvb}_0)$ represents atom positions, atom types, and bond types of the perturbed molecule at time $t$, ground truth molecule, and the predicted molecule respectively, 
${\bm c}(\rvv_t, \rvv_0) = \bm{c}^\star / \sum_{k=1}^{K_a} c_k^\star$ and $\bm{c}^\star(\rvv_t, \rvv_0) = [\alpha_t\rvv_t + (1 - \alpha_t) / K_a] \odot [\bar\alpha_{t-1}\rvv_0 + (1 - \bar\alpha_{t-1}) / K_a]$, 
${\bm c}(\rvb_t, \rvb_0) = \bm{c}^\star / \sum_{k=1}^{K_b} c_k^\star$ and $\bm{c}^\star(\rvb_t, \rvb_0) = [\alpha_t\rvb_t + (1 - \alpha_t) / K_b] \odot [\bar\alpha_{t-1}\rvb_0 + (1 - \bar\alpha_{t-1}) / K_b]$. Note that the condition encoder and the diffusion-based decoder are jointly trained.
}

\revise{
In practice, we set the loss weights as $\gamma_v=100$ and $\gamma_b=100$. Follwing the setting of \citet{pmlr-v202-guan23a}, we set the number of diffusion steps as 1000. For this diffusion noise schedule, we choose to use a sigmoid $\beta$ schedule with $\beta_1 = \texttt{1e-7}$ and $\beta_T = \texttt{2e-3}$ for atom coordinates, and a cosine $\beta$ schedule suggested in \cite{nichol2021improved} with $s=0.01$ for atom types and bond types. 
}

\revise{
We use Adam \cite{kingma2014adam} with \texttt{init\_learning\_rate=0.0005}, \texttt{betas=(0.95, 0.999)} to train the model. And we set \texttt{batch\_size=16} and \texttt{clip\_gradient\_norm=8}. During the training phase, we add a small Gaussian noise with a standard deviation of 0.1 to protein atom coordinates as data augmentation. We also schedule to decay the learning rate exponentially with a factor of 0.6 and a minimum learning rate of 1e-6. The learning rate is decayed if there is no improvement for the validation loss in 10 consecutive evaluations. The evaluation is performed for every 1000 training steps. We trained our model on one NVIDIA GeForce GTX A100 GPU, and it could converge within 237k steps. 
}

\revise{\subsection{Sampling Details}}

\begin{figure*}[h]
\centering
% \raggedleft
% \flushleft
\includegraphics[width=0.98\textwidth]{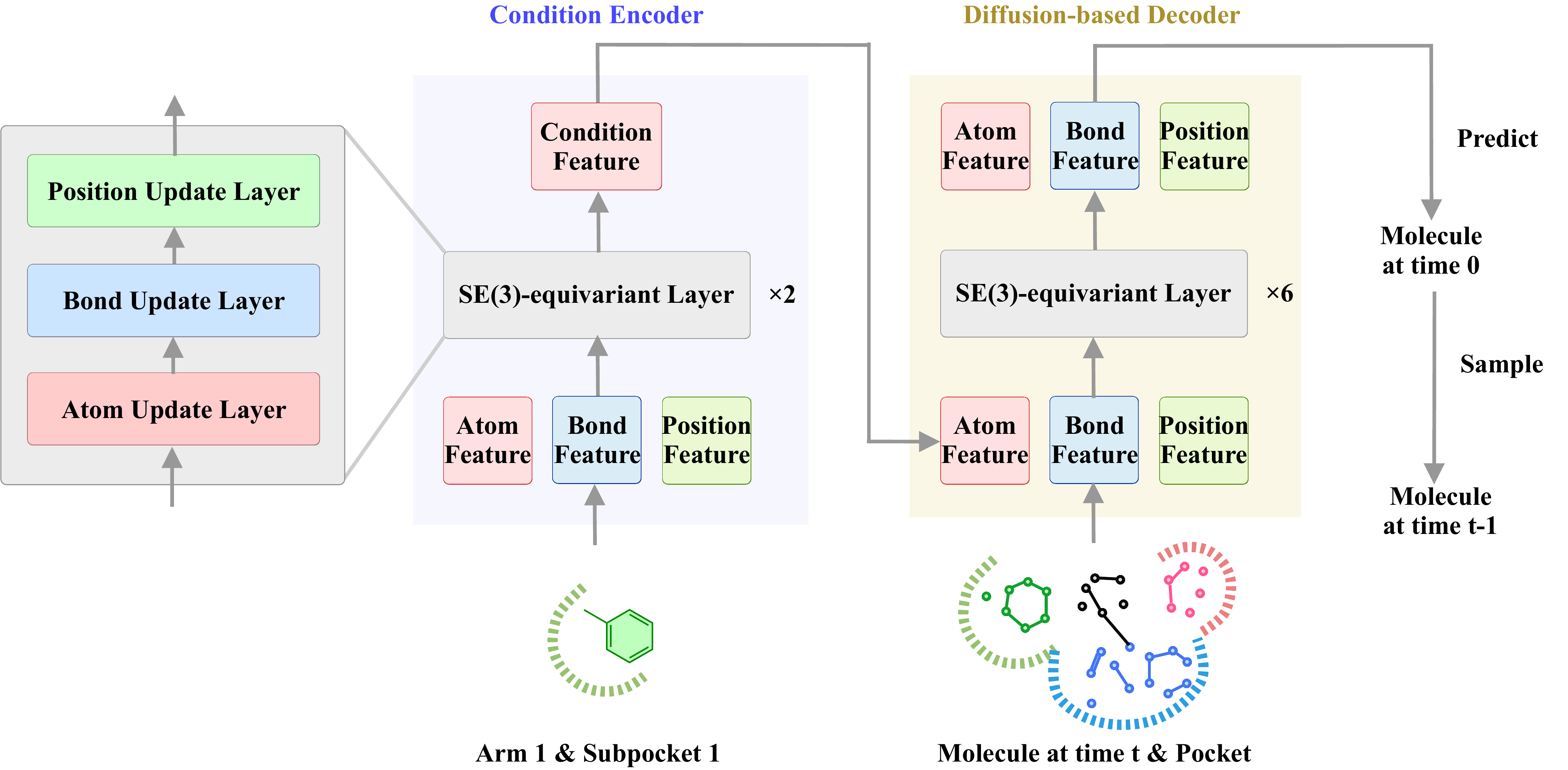}
  \caption{\revise{Illustration of the sampling.}}{\label{fig:sampling_figure_appendix}}
\end{figure*}

\revise{
To sample molecules using the pre-trained controllable and decomposed diffusion model, assume that there are available arms as conditions, we can first sample a noisy molecule from the prior distribution and derive a molecule by iteratively denoising following the reverse process (\eqref{eq:reverse_process}). More specifically, the denoising step at time $t$ corresponds to sampling molecules from the following distributions:
\begin{align}
    & q(\rvx_{t-1}|\rvx_t, \Hat{\rvx}_0)=\gN(\rvx_{t-1};\Tilde{\vmu}_t(\rvx_t,\Hat{\rvx}_0),\Tilde{\beta}_t\rmI), \\
    & q(\rvv_{t-1}|\rvv_t, \Hat{\rvv}_0)=\gC(\rvv_{t-1}|\Tilde{\vc}_{t}(\rvv_t,\Hat{\rvv}_0)), \\
    & q(\rvb_{t-1}|\rvb_t, \Hat{\rvb}_0)=\gC(\rvb_{t-1}|\Tilde{\vc}_{t}(\rvb_t,\Hat{\rvb}_0)),
\end{align}
where $\Tilde{\vmu}_t(\rvx_t,\Hat{\rvx}_0)=\frac{\sqrt{\Bar{\alpha}_{t-1}\beta_t}}{1-\Bar{\alpha}_t}\Hat{\rvx}_0 + \frac{\sqrt{\alpha}_t (1-\Bar{\alpha}_{t-1})}{1-\Bar{\alpha}_{t}} \rvx_t$, $\Tilde{\beta}_t=\frac{1-\Bar{\alpha}_{t-1}}{1-\Bar{\alpha}_{t}}\beta_t$, 
$\Tilde{\bm c}(\rvv_t, \Hat{\rvv}_0) = \Tilde{\bm{c}}^\star / \sum_{k=1}^{K_a} \Tilde{c}_k^\star$ and $\Tilde{\bm{c}}^\star(\rvv_t, \Hat{\rvv}_0) = [\alpha_t\rvv_t + (1 - \alpha_t) / K_a] \odot [\bar\alpha_{t-1}\Hat{\rvv}_0 + (1 - \bar\alpha_{t-1}) / K_a]$, 
$\Tilde{\bm c}(\rvb_t, \Hat{\rvb}_0) = \Tilde{\bm{c}}^\star / \sum_{k=1}^{K_b} \Tilde{c}_k^\star$ and $\Tilde{\bm{c}}^\star(\rvb_t, \Hat{\rvb}_0) = [\alpha_t\rvb_t + (1 - \alpha_t) / K_b] \odot [\bar\alpha_{t-1}\Hat{\rvb}_0 + (1 - \bar\alpha_{t-1}) / K_b]$. Here $(\Hat{\rvx}_0,\Hat{\rvv}_0,\Hat{\rvb}_0)$ is the molecule output by the diffusion-based decoder, whose input is the noisy molecule at time $t$ and the condition feature. The sampling step is illustrated as \cref{fig:sampling_figure_appendix}. During sampling, we also apply validity guidance proposed by \citet{pmlr-v202-guan23a}, which encourages the model to generate molecules with valid structures.}

\revise{\subsection{Optimization Details}}

\begin{figure*}[h]
\centering
% \raggedleft
% \flushleft
\includegraphics[width=0.98\textwidth]{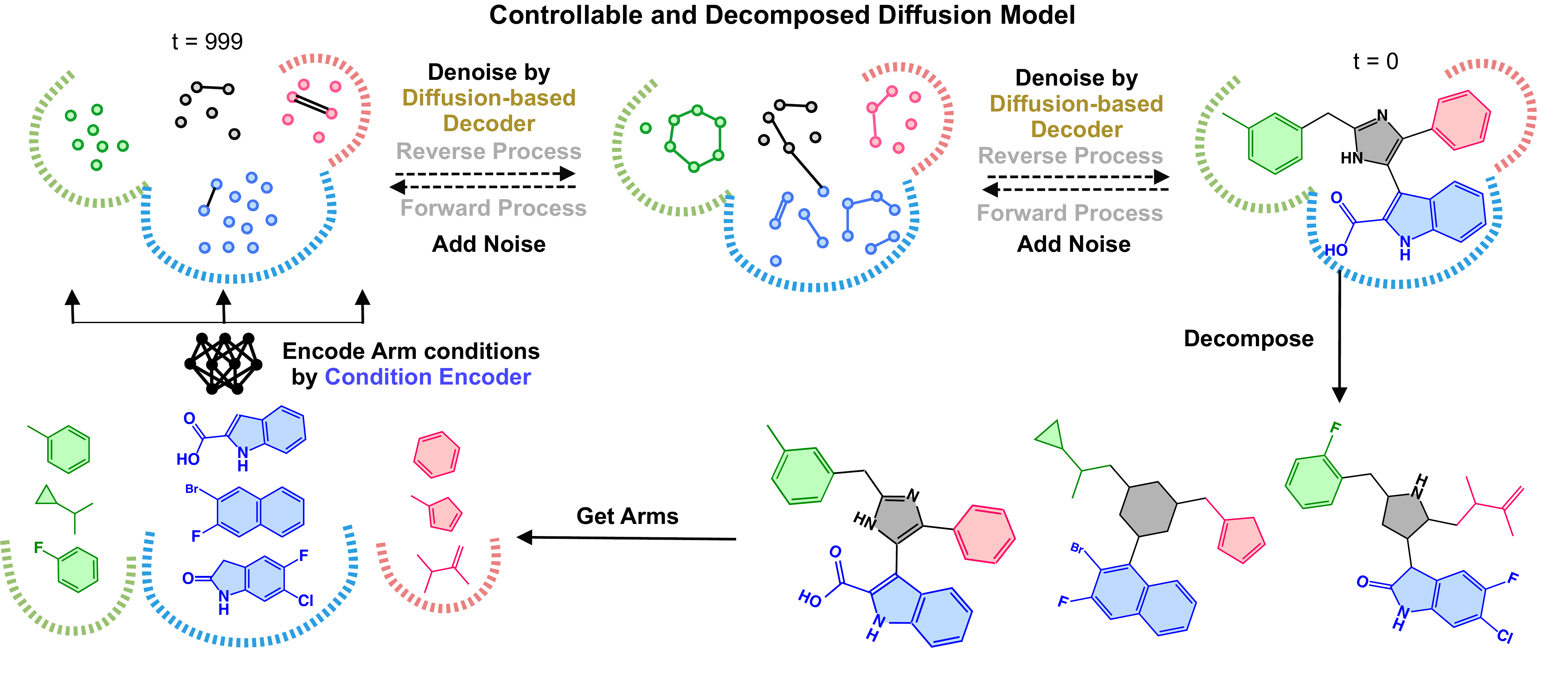}
  \caption{\revise{Illustration of molecular optimization (revised based on \cref{fig:framework}). It is highlighted where we apply the condition encoder and the diffusion-based decoder.}}{\label{fig:overall_framework_appendix}}
\end{figure*}

\revise{
To generate molecules with desired properties, we can apply the pre-trained controllable and decomposed diffusion models for structure-based molecular optimization without any fine-tuning. The optimization procedure is summarized as \cref{alg:optimization} and illustrated as \cref{fig:overall_framework_appendix}. In practice, since reference ligands are not available, we can initialize the arm lists with 20 ligands generated by DecompDiff, and this is the actual setting in our experiment. We have provided the optimization procedure in detail that can be found in \cref{subsec:optimization_procedure} and \cref{subsec:experimental_setup}.
}

\section{Evaluation of Molecular Conformation}
\label{appendix:molecular_conformation}

%  pocket2mol ===
% {'JSD_CC=O': 0.35288391051938683,
%  'JSD_CCC': 0.3231970898708765,
%  'JSD_CCO': 0.40128093814275223,
%  'JSD_CNC': 0.2373004446564458,
%  'JSD_COC': 0.3170516535840215,
%  'JSD_NCC': 0.3506919252052005,
%  'JSD_OPO': 0.2742172919722946}

%  targetdiff ===
% {'JSD_CC=O': 0.3560819015501731,
%  'JSD_CCC': 0.32828224236551484,
%  'JSD_CCO': 0.38511653820579883,
%  'JSD_CNC': 0.3669618993962834,
%  'JSD_COC': 0.3889623741840131,
%  'JSD_NCC': 0.3540600158674388,
%  'JSD_OPO': 0.3028806049929941}

% decompdiff ===
% {'JSD_CC=O': 0.2552716761117448,
%  'JSD_CCC': 0.2936316600898,
%  'JSD_CCO': 0.30849638224934767,
%  'JSD_CNC': 0.3038244903029295,
%  'JSD_COC': 0.32715754087526305,
%  'JSD_NCC': 0.29697460107280155,
%  'JSD_OPO': 0.2527679616134608}

%  decompopt ===
% {'JSD_CC=O': 0.2566049648419848,
%  'JSD_CCC': 0.2802702600511536,
%  'JSD_CCO': 0.33087056304328005,
%  'JSD_CNC': 0.27985379473927297,
%  'JSD_COC': 0.3381760133583762,
%  'JSD_NCC': 0.2664280665200337,
%  'JSD_OPO': 0.19843917914261672}

\begin{figure}[h]
\centering
% \raggedleft
% \flushleft
\includegraphics[width=0.5\textwidth]{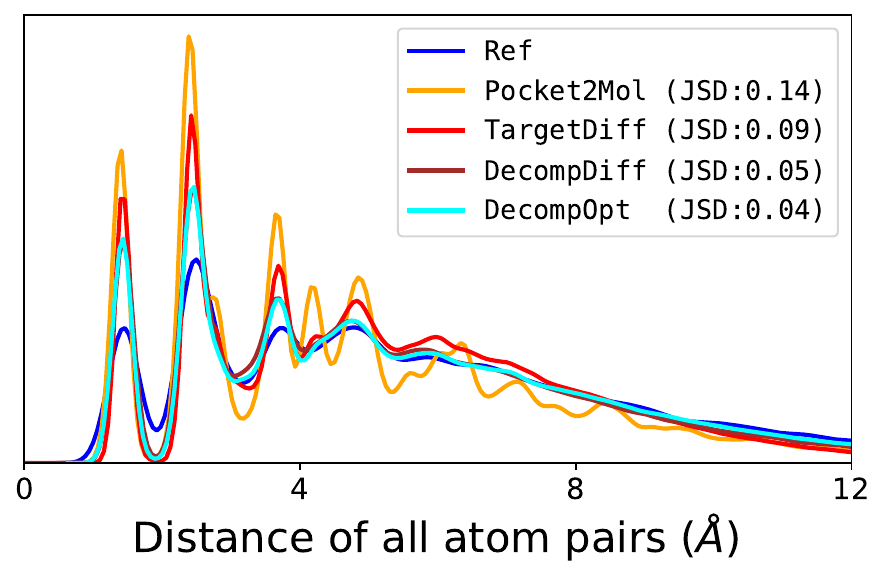}
  \caption{Comparing the distribution for distances of all-atom for reference molecules in the test set and model-generated molecules. Jensen-Shannon divergence (JSD) between two distributions is reported.}{\label{fig:atom_distance_distribution}}
\end{figure}

To evaluate generated molecules from the perspective of molecular conformation, we compute the Jensen-Shannon divergences (JSD) in atom distance distributions between the reference molecules and the generated molecules (see \cref{fig:atom_distance_distribution}). 
% bond distance left

We also compute different bond distance and bond angle distributions of the generated molecules and compare them against the corresponding reference empirical distributions in \cref{tab:bond_jsd,tab:bond_angle_jsd}.
% , and plot the carbon-carbon bond distance distribution and all-atom pairwise distance distribution of the generated molecules in \cref{fig:atom_jsd}. 

\begin{table}[h]
    \centering
    \caption{Jensen-Shannon divergence between bond distance distributions of the reference molecules and the generated molecules, and lower values indicate better performances. ``-'', ``='', and ``:'' represent single, double, and aromatic bonds, respectively. We highlight the best two results with \textbf{bold text} and \underline{underlined text}, respectively.}
    % \small
    % \begin{adjustbox}{width=1.0\textwidth}
        \begin{tabular}{cccccccc}
\toprule
{Bond} & liGAN & GraphBP & AR & \thead{Pocket2 \\ Mol} & \thead{Target \\ Diff}& \thead{Decomp \\ Diff} & Ours \\
\midrule
C$-$C & 0.601 & 0.368 & 0.609 & 0.496 & {0.369} & \textbf{0.359} & \underline{0.362}\\
C$=$C & 0.665 & 0.530 & 0.620 & 0.561 & \underline{0.505} & {0.537} & \textbf{0.504}\\
C$-$N & 0.634 & 0.456 & 0.474 & 0.416 & {0.363} & \underline{0.344} & \textbf{0.328}\\
C$=$N & 0.749 & 0.693 & 0.635 & 0.629 & \underline{0.550} & {0.584} & \textbf{0.566}\\
C$-$O & 0.656 & 0.467 & 0.492 & 0.454 & 0.421 & \underline{0.376}& \textbf{0.373} \\
C$=$O & 0.661 & 0.471 & 0.558 & 0.516 & {0.461} & \underline{0.374} & \textbf{0.329}\\
C$:$C & 0.497 & 0.407 & 0.451 & 0.416 & {0.263} & \underline{0.251}& \textbf{0.196}\\
C$:$N & 0.638 & 0.689 & 0.552 & 0.487 & \underline{0.235} & {0.269} & \textbf{0.219}\\
\bottomrule
\end{tabular}
        \label{tab:bond_jsd}
    % \end{adjustbox}
\end{table}

\begin{table}[h]
    \centering
    \caption{Jensen-Shannon divergence between bond angle distributions of the reference molecules and the generated molecules, and lower values indicate better performances. We highlight the best two results with \textbf{bold text} and \underline{underlined text}, respectively.}
    % \small
    % \begin{adjustbox}{width=0.5\textwidth}
        \begin{tabular}{cccccccc}
\toprule
{Angle} & liGAN & GraphBP & AR & \thead{Pocket2 \\ Mol} & \thead{Target \\ Diff} & \thead{Decomp \\ Diff}  & Ours \\
\midrule
CCC & 0.598 & 0.424 & 0.340 & 0.323 & 0.328 & \underline{0.314} & \textbf{0.280} \\
CCO & 0.637 & {0.354} & 0.442 & 0.401 & 0.385 & \underline{0.324} & \textbf{0.331}  \\ 
CNC & 0.604 & 0.469 & 0.419 & \textbf{0.237} & 0.367 & {0.297} &  \underline{0.280}\\
OPO & 0.512	& 0.684 & 0.367	& {0.274}	& 0.303	& \underline{0.217} &  \textbf{0.198}\\
NCC & 0.621	& 0.372	& 0.392	& 0.351	& 0.354	& \underline{0.294} & \textbf{0.266} \\
CC=O & 0.636 & 0.377 & 0.476 & 0.353 & 0.356 & \underline{0.259} & \textbf{0.257} \\
COC  & 0.606 & 0.482 & 0.459 & \textbf{0.317} & 0.389 & {0.339} & \underline{0.338} \\
\bottomrule
\end{tabular}
\label{tab:bond_angle_jsd} 
    % \end{adjustbox}
\end{table}

\begin{figure}[H]
\centering
\includegraphics[width=0.8\textwidth]{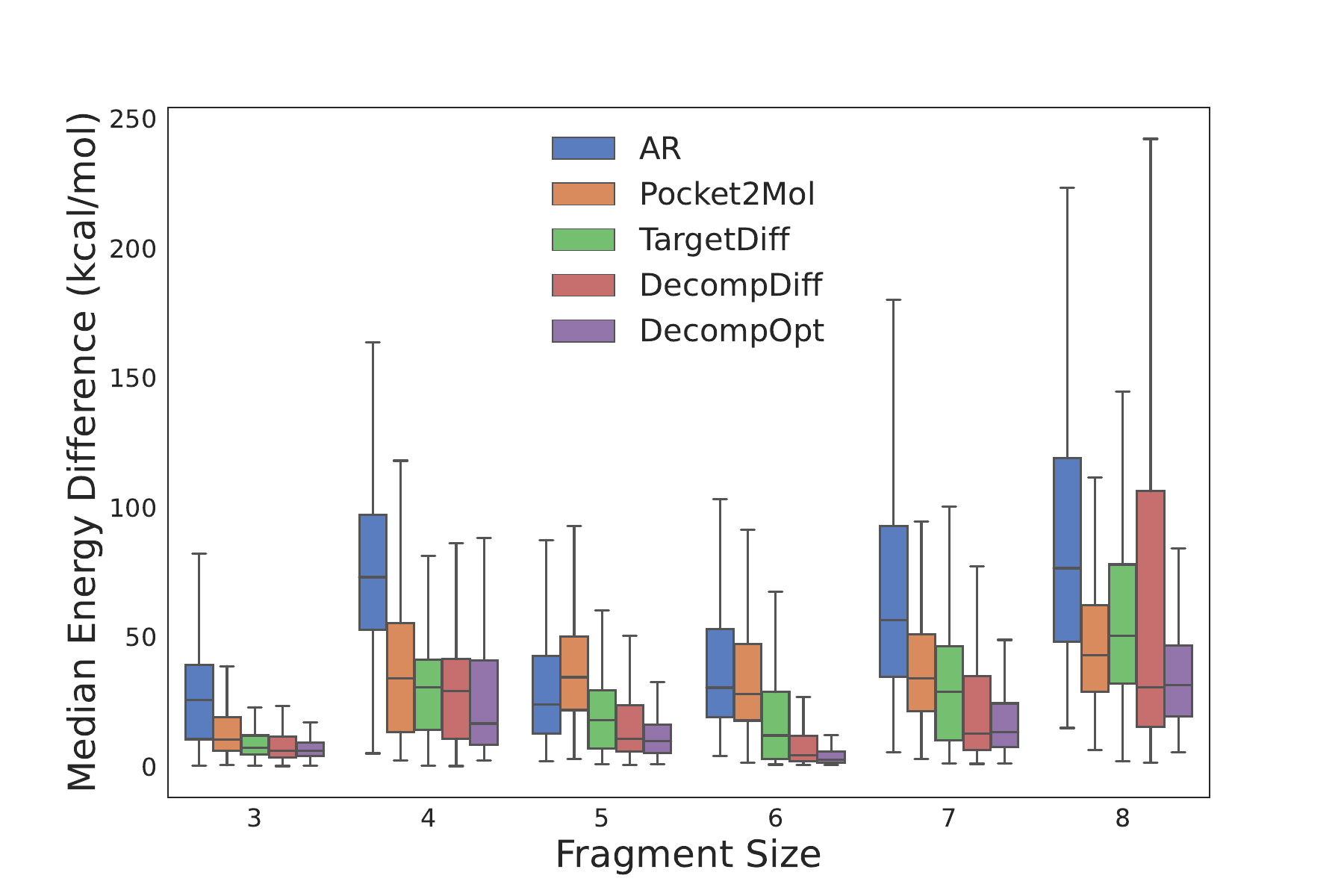}
  \caption{\revise{Median energy difference for molecules with different number of rotatable bonds before and after the force-field optimization.}}{\label{fig:energy_diff}}
\end{figure}

\begin{figure}[h]
\centering
% \raggedleft
% \flushleft
\includegraphics[width=0.8\textwidth]{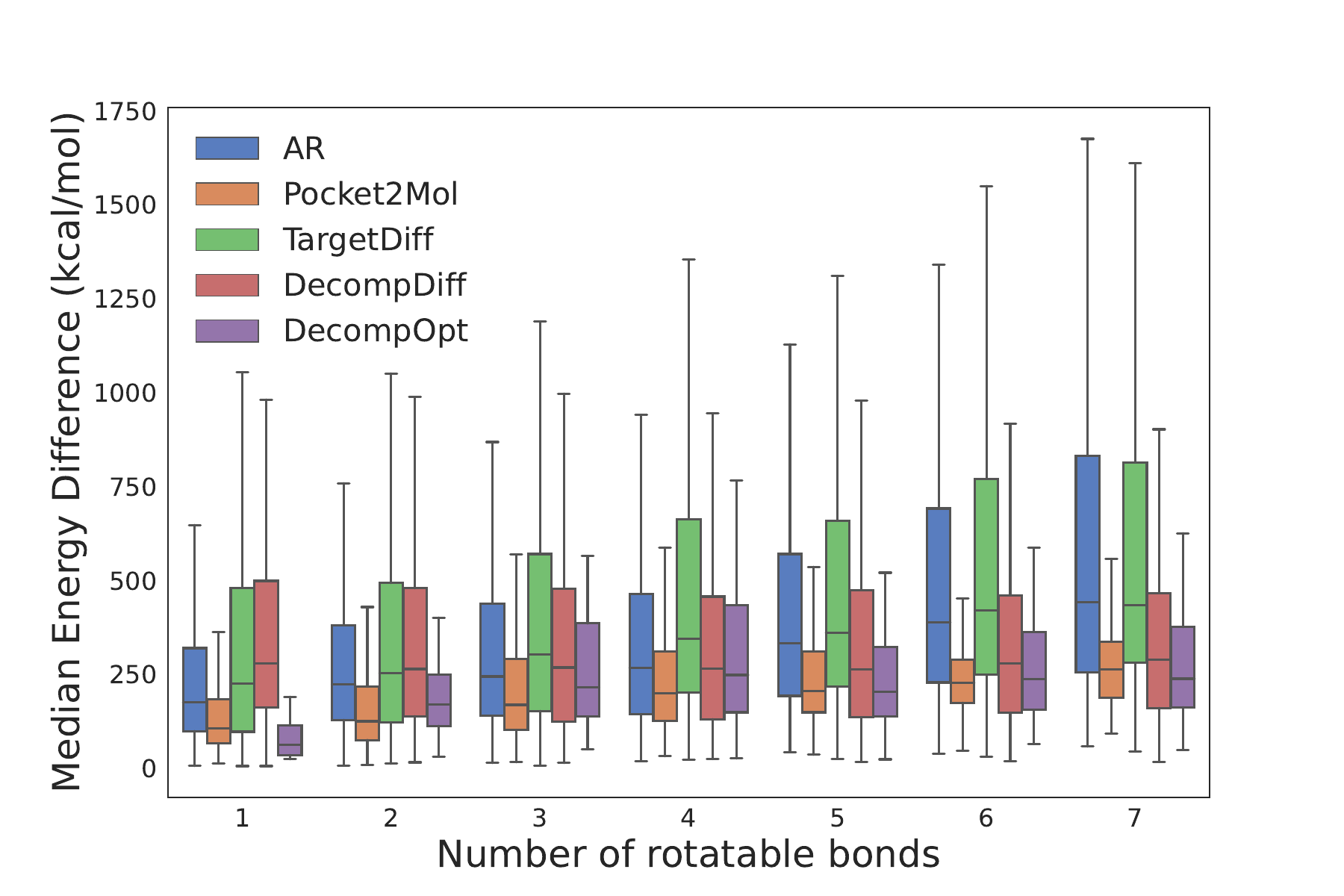}
  \caption{\revise{Median energy difference for molecules with different number of rotatable bonds before and after the force-field optimization.}}{\label{fig:energy_diff_whole}}
\end{figure}

\revise{To further measure the quality of generated conformation, we optimize the generated structures with Merck Molecular Force Field (MMFF) \citep{halgren1996merck} and calculate the energy difference between pre- and pos- MMFF-optimized coordinates for different rigid fragments that do not contain any rotatable bonds. As \cref{tab:energy_diff_frag} and \cref{fig:energy_diff} show, \method achieves low energy differences and outperforms baselines in most cases. We also calculate the energy difference before and after force field optimization for the whole molecules. As \cref{tab:energy_diff_whole} and \cref{fig:energy_diff_whole} show, notably, \method outperforms all diffusion-based methods by a large margin and achieve comparable performance with the best baseline. These results show that the conformation of ligands generated by \method is high-quality and stable.} 

\begin{table}[H]
    \centering
    \caption{\revise{Median energy difference for rigid fragment of different fragment size (3/4/5/6/7/8 atoms) before and after the force-field optimization.}}
    % \small
    % \begin{adjustbox}{width=0.5\textwidth}
        \begin{tabular}{c|cccccc}
\toprule
\multirow{2}{*}{Methods} & \multicolumn{6}{c}{Median Energy Difference ($\downarrow$)} \\
& 3 & 4 & 5 & 6 & 7 & 8 \\
\midrule
LiGAN & 86.32 & 165.15 & 105.96 & 185.70 & 243.79 & 332.81 \\
AR & 25.79 & 73.06 & 23.89 & 30.42 & 56.47 & 76.50 \\ 
Pocket2Mol & 10.43 & 33.93 & 34.47 & 27.86 & 33.90 & 42.97 \\
TargetDiff & 7.31 & 30.57 & 18.01 & 11.98 & 28.92 & 50.42 \\
DecompDiff & 6.01 & 29.20 & 10.78 & 4.33 & 12.74 & 30.68 \\
\method & 6.00 & 16.59 & 9.89 & 2.61 & 13.29 & 31.49 \\
\bottomrule
\end{tabular}
\label{tab:energy_diff_frag} 
    % \end{adjustbox}
\end{table}

\begin{table}[h]
    \centering
    \caption{\revise{Median energy difference for molecules with different number of rotatable bonds (1/2/3/4/5/6/7 rotatable bonds) before and after the force-field optimization.}}
    % \small
    % \begin{adjustbox}{width=0.5\textwidth}
        \begin{tabular}{c|ccccccc}
\toprule
\multirow{2}{*}{Methods} & \multicolumn{7}{c}{Median Energy Difference ($\downarrow$)} \\
& 1 & 2 & 3 & 4 & 5 & 6 & 7 \\
\midrule
LiGAN & 810.45 & 981.53 & 1145.96 & 1783.95 & 1960.24 & 2547.32 & 2735.75 \\
AR & 176.67 & 222.74 & 244.51 & 268.01 & 332.89 & 388.70 & 441.90 \\ 
Pocket2Mol & 105.64 & 125.19 & 168.84 & 199.33 & 204.82 & 226.73 & 263.96 \\
TargetDiff & 225.48 & 253.72 & 303.60 & 344.12 & 360.74 & 420.47 & 434.30 \\
DecompDiff & 279.44 & 264.16 & 268.23 & 265.57 & 262.69 & 279.73 & 289.07 \\
\method & 63.33 & 169.17 & 215.19 & 248.35 & 202.81 & 237.38 & 238.32 \\
\bottomrule
\end{tabular}
\label{tab:energy_diff_whole} 
    % \end{adjustbox}
\end{table}

\section{Additional Results}\label{appendix:additional_results}

\begin{figure}[H]
\centering
% \raggedleft
% \flushleft
\includegraphics[width=1.0\textwidth]{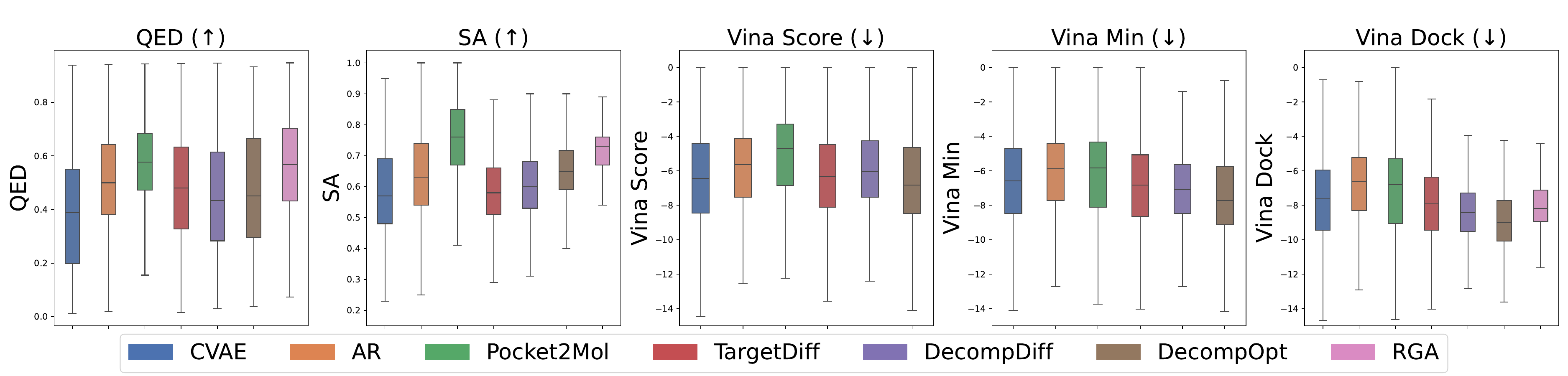}
  \caption{\revise{The boxplots of QED, SA, Vina Score, Vina Minimize, and Vina Dock of ligands generated by \method and baseline models.}}{\label{fig:metrics_boxplot}}
\end{figure}

\revise{\subsection{Full evaluation results}}
\revise{We provide box plots of evaluation metrics as shown in \cref{fig:metrics_boxplot}.}

Following \citet{pmlr-v202-guan23a}, our model also has variants of priors. 
\cref{tab:main_tab_full} shows the results of multiple variants of our models. The setting of \textit{Ref Prior}, \textit{Pocket Prior}, and \textit{Opt Prior} strictly follows DecompDiff \citep{pmlr-v202-guan23a}. \textit{Ref Best} means using the best checkpoint instead of the last checkpoint for each target pocket during optimization with reference priors for evaluation. For \textit{Pocket/Opt Best}, it is similar. \textit{Best of Best} means using the best checkpoint across all checkpoints with \textit{Ref Prior} and \textit{Pocket Prior} during optimization for each target pocket.

\begin{table}[h]
    \centering
    \caption{Summary of different properties of reference molecules and molecules generated by our model and other generation (Gen.) and optimization (Opt.) baselines. ($\uparrow$) / ($\downarrow$) denotes a larger / smaller number is better.
    }
    \begin{adjustbox}{width=1\textwidth}
    \renewcommand{\arraystretch}{1.2}
    \begin{tabular}{c|cc|cc|cc|cc|cc|cc|cc|c}
    \toprule
    % \diagbox{Model}{Metric} 
    \multicolumn{1}{c|}{\multirow{2}{*}{Methods}} & \multicolumn{2}{c|}{Vina Score ($\downarrow$)} & \multicolumn{2}{c|}{Vina Min ($\downarrow$)} & \multicolumn{2}{c|}{Vina Dock ($\downarrow$)} & \multicolumn{2}{c|}{High Affinity ($\uparrow$)} & \multicolumn{2}{c|}{QED ($\uparrow$)}   & \multicolumn{2}{c|}{SA ($\uparrow$)} & \multicolumn{2}{c|}{Diversity ($\uparrow$)} & \multicolumn{1}{c}{Success} \\
    \multicolumn{1}{c|}{} & Avg. & Med. & Avg. & Med. & Avg. & Med. & Avg. & Med. & Avg. & Med. & Avg. & Med. & Avg. & Med. & Rate \\
    
    \toprule
    
    \multicolumn{1}{c|}{Reference}   & -6.36 & -6.46 & -6.71 & -6.49 & -7.45 & -7.26 & -  & - & 0.48 & 0.47 & 0.73 & 0.74 & - & - & 25.0\%  \\

   \midrule

    \method (Ref Prior) & -5.68 & -5.88 & -6.53  & -6.49 &  -7.49 & -7.66 & 59.2\% & 65.0\% & 0.56 & 0.58 & 0.73 & 0.73 & 0.64 & 0.66 & 35.4\% \\

    \method (Ref Best) & -5.75 & -5.97 & -6.58 & -6.70 & -7.63 & -8.02 & 62.6\% & 74.3\% & 0.56 & 0.59 & 0.73 & 0.72 & 0.63 & 0.67 & 39.4\% \\
    
    \method (Pocket Prior) & -5.27 & -6.38 & -7.07 & -7.45 & -8.85 & -8.72 & 71.4\% & 93.8\% & 0.40 & 0.36 & 0.63 & 0.63 & 0.60 & 0.61 & 29.2\% \\

    \method (Pocket Best) & -5.33 & -6.49 & -7.08 & -7.60 & -9.01 & -8.98 & 73.9\% & 100\% & 0.41 & 0.39 & 0.63 & 0.63 & 0.59 & 0.60 & 44.7\% \\
    
    \method (Opt Prior) & -5.73  & -6.64 & -7.29 & -7.53 & -8.78 & -8.72 & 70.3\% & 89.9\% & 0.46 & 0.44 & 0.65 & 0.65 & 0.61 & 0.61 & 38.1\% \\

    \method (Opt Best) & -5.87  & -6.81 & -7.35 & -7.72 & -8.98 & -9.01 & 73.5\% & 93.3\% & 0.48 & 0.45 & 0.65 & 0.65 & 0.60 & 0.61 & 52.5\% \\

    \method (Best of Best) & -6.22  & -6.94 & -7.50 & -7.74 & -8.98 & -8.95 & 76.2\% & 100\% & 0.51 & 0.51 & 0.67 & 0.67 & 0.61 & 0.63 & 60.6\% \\
    \bottomrule
    \end{tabular}
    \renewcommand{\arraystretch}{1}
    \end{adjustbox}\label{tab:main_tab_full}
\end{table}

\begin{figure}[h]
\centering
% \raggedleft
% \flushleft
\includegraphics[width=0.7\textwidth]{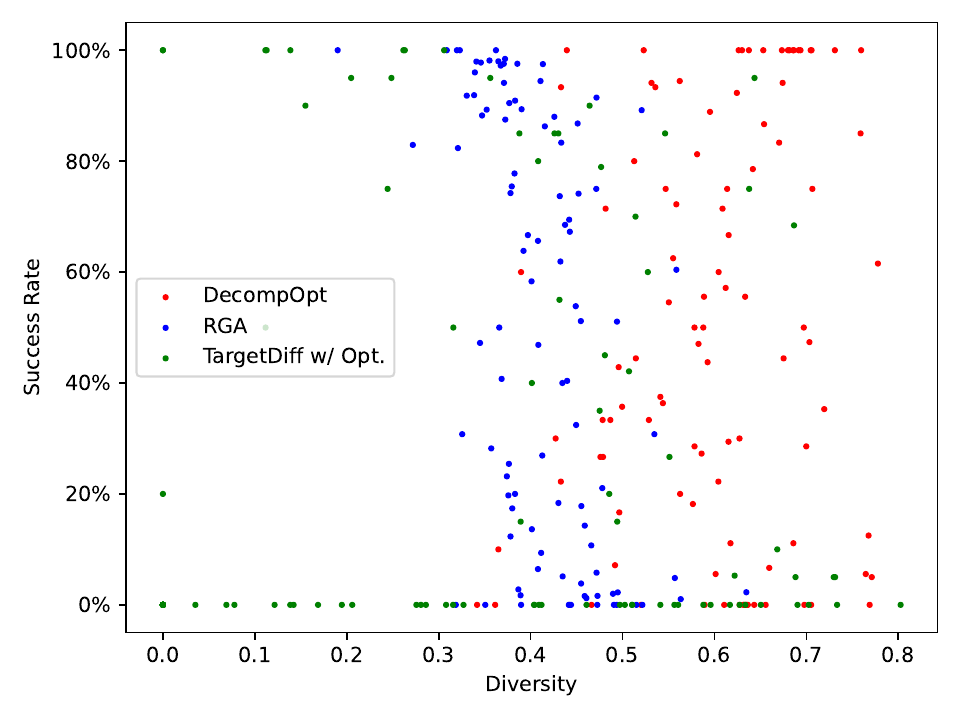}
  \caption{Trade-off of Success Rate and diversity. Each point with coordinate $(x,y)$ represents a pocket with Success Rate $x$ and diversity $y$. The closer to the top right, the better.}{\label{fig:diversity_scatter}}
\end{figure}

\revise{\subsection{Trade-off between Success Rate and Diversity}}
In addition to overall performance, we also show the trade-off between Success Rate and diversity of RGA, TargetDiff w/ Opt., and \method for each target protein pocket in \cref{fig:diversity_scatter}. \method shows general superiority to the other two baselines in most cases considering both Success Rate and diversity. 

\revise{\subsection{Evaluation of the ability to design novel ligands}}

\revise{We additionally test the \textbf{Novelty} and \textbf{Similarity} of generated ligands compared with the reference ligand. Novelty is defined as the ratio of generated ligands that are different from the reference ligands of the corresponding pockets in the test set. Similarity is defined as the Tanimoto Similarity between the generated ligands and the corresponding reference ligands. The results show that the generated ligands are not similar to reference ligands in the test set. Besides, we also test \textbf{Uniqueness} and \textbf{Diversity} of generated ligands. Uniqueness is the percentage of unique molecules among all the generated molecules. Diversity is the same as that in \cref{subsec:experimental_setup}. The results are reported in \cref{tab:similarity}. These results show that \method can design novel ligands, which is an important ability for drug discovery.}

\begin{table}[H]
    \centering
    \caption{\revise{Evaluation of the ability to design novel ligands.}}
    % \small
    % \begin{adjustbox}{width=0.5\textwidth}
        \begin{tabular}{c|cccc}
\toprule
{Methods} & Novelty & Similarity & Uniqueness & Diversity \\
\midrule
LiGAN & 100$\%$ & 0.22 & 87.82$\%$ & 0.66\\ 
AR &  100$\%$ & 0.24  & 100$\%$ & 0.70\\ 
Pocket2Mol &  100$\%$ & 0.26 & 100$\%$ & 0.69\\
TargetDiff &  100$\%$ & 0.30 & 99.63$\%$ & 0.72\\
DecompDiff &  100$\%$ & 0.34 & 99.99$\%$ & 0.68\\
RGA & 100$\%$ & 0.37 & 96.82$\%$ & 0.41\\
\method & 100$\%$ & 0.36 & 100$\%$ & 0.60\\
\bottomrule
\end{tabular}
\label{tab:similarity} 
    % \end{adjustbox}
\end{table}

\revise{\subsection{Effects of subpockets.}}

\revise{To study the influence of subpockets in controlling the optimization, we further conducted an ablation study using only arms without subpockets as conditions. As \cref{tab:effect_of_subpocket} shows, while \method, when solely with arms as conditions, is capable of optimizing all metrics, its efficiency in this scenario is not as well as \method that utilizes both arms and pockets as conditions. Recall that we use SE(3)-invaraint features of arms (and subpockets) as conditions. Without subpockets, this feature would be agnostic to the molecular interaction and spatial relation between the arms and subpockets. Such information is important to some of the properties (e.g., Vina scores). The SE(3)-invaraint features from pairs of subpockets and arms contain the aforementioned information and are better aligned with the protein-ligand complex being generated.}

\begin{table}[h]
    \centering
    \caption{\revise{Comparison of \method optimization results with only arms and arm-pocket complexes as conditions. ($\uparrow$) / ($\downarrow$) denotes a larger / smaller number is better.}
    }
    \begin{adjustbox}{width=1\textwidth}
    \renewcommand{\arraystretch}{1.2}
    \begin{tabular}{c|cc|cc|cc|cc|cc|cc|cc|c}
    \toprule
    % \diagbox{Model}{Metric} 
    \multicolumn{1}{c|}{\multirow{2}{*}{Methods}} & \multicolumn{2}{c|}{Vina Score ($\downarrow$)} & \multicolumn{2}{c|}{Vina Min ($\downarrow$)} & \multicolumn{2}{c|}{Vina Dock ($\downarrow$)} & \multicolumn{2}{c|}{High Affinity ($\uparrow$)} & \multicolumn{2}{c|}{QED ($\uparrow$)}   & \multicolumn{2}{c|}{SA ($\uparrow$)} & \multicolumn{2}{c|}{Diversity ($\uparrow$)} & \multicolumn{1}{c}{Success} \\
    \multicolumn{1}{c|}{} & Avg. & Med. & Avg. & Med. & Avg. & Med. & Avg. & Med. & Avg. & Med. & Avg. & Med. & Avg. & Med. & Rate \\
    
    \toprule

    DecompDiff & -5.67 & -6.04 & -7.04 & -7.09 & -8.39 & -8.43 & 64.4\% & 71.0\% & 0.45 & 0.43 & 0.61 & 0.60 & 0.68 & 0.68 & 24.5\% \\

    \method (arms-only) & -5.52 & -6.26 & -7.05 & -7.26 & -8.65 & -8.64 & 66.6\% & 86.1\% & 0.46 & 0.43 & 0.63 & 0.63 & 0.63 & 0.63 & 45.7\% \\
    
    \method & -5.87  & -6.81 & -7.35 & -7.72 & -8.98 & -9.01 & 73.5\% & 93.3\% & 0.48 & 0.45 & 0.65 & 0.65 & 0.60 & 0.61 & 52.5\% \\
    \bottomrule
    \end{tabular}
    \renewcommand{\arraystretch}{1}
    \end{adjustbox}\label{tab:effect_of_subpocket}
\end{table}

\revise{\subsection{Influence of the quality of initial ligands on performance.} 
\revise{To study the influence of the quality of initial ligands on performance of structure-based molecular optimization, we have conducted an ablation study focusing on Vina Min score optimization, using ligands with high and low Vina Min scores as initializations for the arm lists.
Due to limited resources, we chose to conduct this study on the protein 2V3R, which is randomly chosen from our test set. We generated 100 ligands using DecompDiff and selected 20 ligands with the highest and lowest Vina Min scores. These ligands were then used as the initial conditions for the optimization process. As shown in \cref{tab:initial_ligand_influence}, the optimization outcomes are slightly influenced by the quality of the initial ligands. However, regardless the quality of the initial ligands, \method can consistently improve the quality of the generated ligands.}}

\begin{table}[H]
    \centering
    \caption{\revise{Comparison of optimization with initial ligands of different quality.}}
    % \small
    % \begin{adjustbox}{width=0.5\textwidth}
        \begin{tabular}{c|cc|cc|c}
\toprule
\multicolumn{1}{c|}{\multirow{2}{*}{}} & \multicolumn{2}{c|}{\multirow{1}{*}{High Vina Min Scores}} & \multicolumn{2}{c|}{\multirow{1}{*}{Low Vina Min Scores}} &  \multicolumn{1}{c}{\multirow{1}{*}{$\Delta$ (high - low)}}    \\
 \multicolumn{1}{c|}{} & Avg. & Med. & Avg. & Med. & Avg. \\
\midrule
Initial ligands & -8.54 & -8.48 & -7.08 & -7.04 & -1.46 \\ 
\method &  -9.12 & -8.96  & -9.00 & -8.96 & -0.12 \\ 
\bottomrule
\end{tabular}
\label{tab:initial_ligand_influence} 
    % \end{adjustbox}
\end{table}

\revise{\subsection{Influence of the number of initial ligands on performance.}} 

\revise{To study the influence of the number of initial molecules on the performance of structure-based molecular optimization, 
we further run the experiments with initial arm lists of 1 and 5 molecules generated by DecompDiff. As \cref{tab:ablation_init_ligand_num} indicates, the initial number of molecules has a modest impact on the optimization outcomes, with a higher number of molecules generally leading to improved performance. Notably, even when starting with a single molecule generated by DecompDiff, \method demonstrates a considerably high success rate.}

\begin{table}[h]
    \centering
    \caption{\revise{Summary of results using different number of molecules to initialize arm lists. ($\uparrow$) / ($\downarrow$) denotes a larger / smaller number is better.}
    }
    \begin{adjustbox}{width=1\textwidth}
    \renewcommand{\arraystretch}{1.2}
    \begin{tabular}{c|cc|cc|cc|cc|cc|cc|cc|c}
    \toprule
    % \diagbox{Model}{Metric} 
    \multicolumn{1}{c|}{\multirow{2}{*}{Methods}} & \multicolumn{2}{c|}{Vina Score ($\downarrow$)} & \multicolumn{2}{c|}{Vina Min ($\downarrow$)} & \multicolumn{2}{c|}{Vina Dock ($\downarrow$)} & \multicolumn{2}{c|}{High Affinity ($\uparrow$)} & \multicolumn{2}{c|}{QED ($\uparrow$)}   & \multicolumn{2}{c|}{SA ($\uparrow$)} & \multicolumn{2}{c|}{Diversity ($\uparrow$)} & \multicolumn{1}{c}{Success} \\
    \multicolumn{1}{c|}{} & Avg. & Med. & Avg. & Med. & Avg. & Med. & Avg. & Med. & Avg. & Med. & Avg. & Med. & Avg. & Med. & Rate \\
    
    \toprule

    init num = 1 & -5.41 & -6.61 & -7.12 & -7.51 &  -8.78 & -8.82 & 70.4\% & 88.9\% & 0.47 & 0.45 & 0.64 & 0.63 & 0.61 & 0.61 & 47.0\% \\

    init num = 5 & -5.71 & -6.71 & -7.25 & -7.58 & -8.86 & -8.97 & 71.8\% & 93.3\% & 0.49 & 0.46 & 0.65 & 0.64 & 0.60 & 0.61 & 49.4\% \\
    
    init num = 20 & -5.87  & -6.81 & -7.35 & -7.72 & -8.98 & -9.01 & 73.5\% & 93.3\% & 0.48 & 0.45 & 0.65 & 0.65 & 0.60 & 0.61 & 52.5\% \\
    \bottomrule
    \end{tabular}
    \renewcommand{\arraystretch}{1}
    \end{adjustbox}\label{tab:ablation_init_ligand_num}
\end{table}

\section{Extended Results of Controllability}
\label{appendix:controllability}

\subsection{R-group Optimization}

\begin{figure}[H]
\centering
% \raggedleft
% \flushleft
\includegraphics[width=0.8\textwidth]{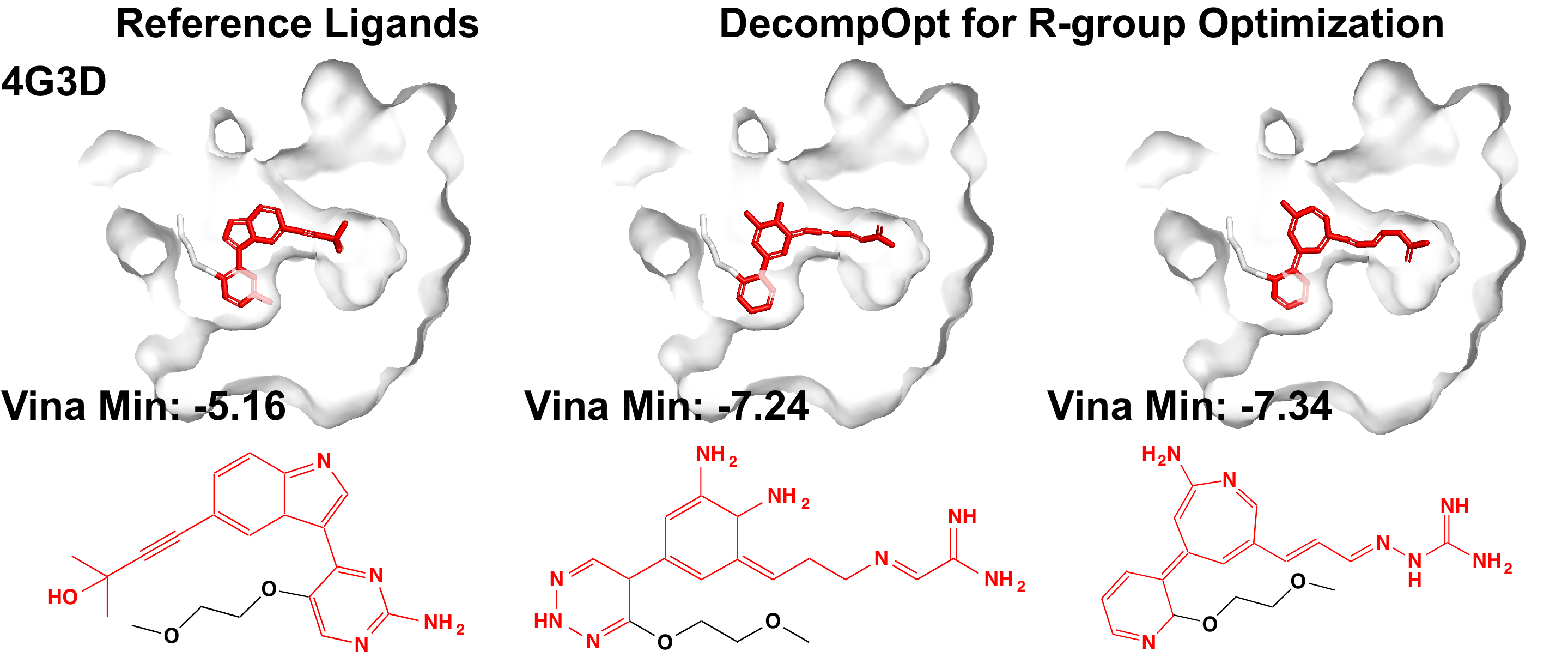}
  \caption{\revise{Additional R-group optimization result. The left column is reference binding molecule, the middle and right columns are molecules generated by \method with 30 rounds of optimization on protein 4G3D. Optimized R-group are highlighted in red.}}{\label{fig:additonal_r_group}}
\end{figure}

\revise{We provide additional R-group Optimization experiment on protein 4G3D, as shown in \cref{fig:additonal_r_group}.}

% \begin{center}
%   \includegraphics[width=0.8\textwidth]{figures/additional_r-group.pdf}
%   \captionof{figure}{\revise{Additional R-group optimization result. The left column is reference binding molecule, the middle and right columns are molecules generated by \method with 30 rounds of optimization on protein 4G3D. Optimized R-group are highlighted in red.}}
%   {\label{fig:additonal_r_group}}
% \end{center}

\begin{table}[H]
    \centering
    \caption{R-group optimization results generated using Decompdiff and \method on protein 3DAF and 4F1M. \method was optimized over 30 rounds towards high \revise{Vina Min Score} and evaluated using the final round results. Both targets were assessed with 20 generated molecules and the mean of properties are reported.}
    \begin{adjustbox}{width=1\textwidth}
    \renewcommand{\arraystretch}{1.2}
    \begin{tabular}{l|ccc|ccc}
    \toprule
    % \diagbox{Model}{Metric} 
    \multicolumn{1}{c|}{\multirow{2}{*}{Model}} & 
    \multicolumn{3}{c|}{3DAF} & 
    \multicolumn{3}{c}{4F1M} \\
    \multicolumn{1}{c|}{} & Vian Min ($\downarrow$) & Tanimoto Sim. ($\uparrow$) & Complete. ($\uparrow$) & Vian Min ($\downarrow$) & Tanimoto Sim. ($\uparrow$) & Complete. ($\uparrow$) \\
    
    \toprule
    Decompdiff & -8.44 & 0.15 & 60.0\% & -5.90 & 0.15 & 65.0\% \\
    
    \method & -9.39 & 0.23 & 95.0\% & -6.32 & 0.49 & 55.0\% \\
    
    \bottomrule
    \end{tabular}
    \renewcommand{\arraystretch}{1}
    \end{adjustbox}\label{tab:r-group_eval}
\end{table}

\subsection{Fragment Growing}
\begin{figure}[H]
\centering
% \raggedleft
% \flushleft
\includegraphics[width=0.8\textwidth]{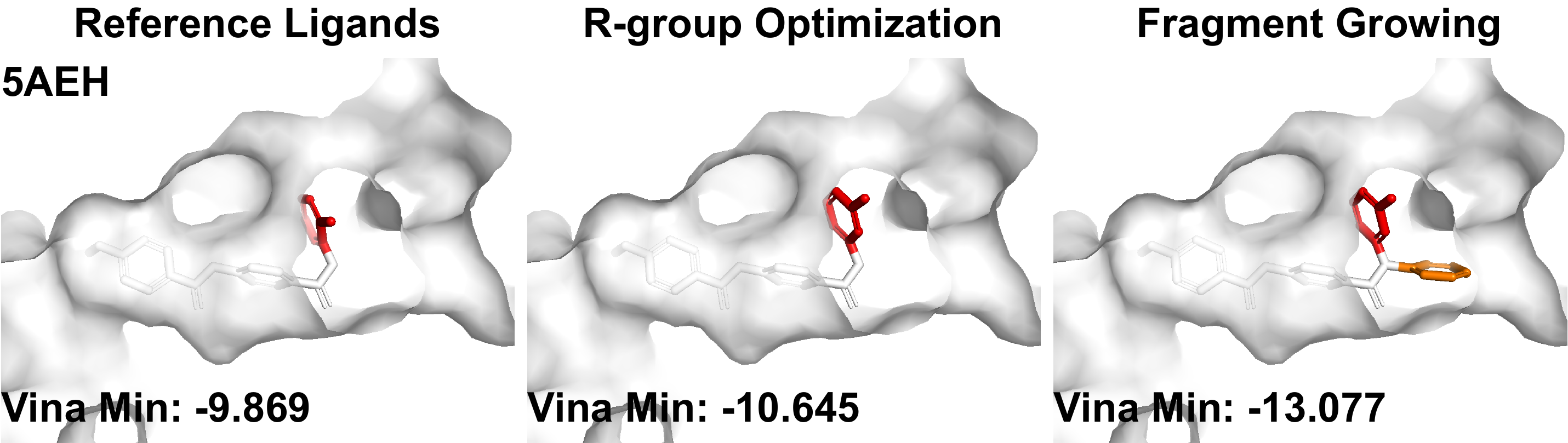}
  \caption{Example of R-group optimization and fragment growing conducted using \method on 5AEH. The reference ligand, the best R-group result, and the best fragment growing result based on R-group optimization are displayed from left to right. The selected R-group is highlighted in red, while the newly extended arm is highlighted in orange.}{\label{fig:fragment_growing}}
\end{figure}

Enhancing the binding affinity of drug candidates through combination of R-group optimization and fragment growing can effectively leverage capabilities of \method. The quantitative results are shown in \cref{tab:r-group_eval}. For our case study, we perform R-group optimization and fragment growing on 5AEH. Starting from a high binding affinity drug candidate, we first optimize R-group for 30 rounds same as workflow in Section 4. Subsequently, we design the new arms prior and atom num with expert guidance, and expand fragments using \method. As \cref{fig:fragment_growing} shows, \method ultimately generates molecules with a \revise{Vina Min Score} more than 4 kcal/mol
better than the reference.

% \begin{center}
%   \includegraphics[width=0.8\textwidth]{figures/fragment_grow_update.pdf}
%   \captionof{figure}{Example of R-group optimization and fragment growing conducted using \method on 5AEH. The reference ligand, the best R-group result, and the best fragment growing result based on R-group optimization are displayed from left to right. The selected R-group is highlighted in red, while the newly extended arm is highlighted in orange.}
%   {\label{fig:fragment_growing}}
% \end{center}

\subsection{Scaffold Hopping}

\paragraph{Additional Evaluation Metrics}

In addition to evaluation metrics discussed in Section 4, we evaluated \textit{Validity}, \textit{Uniqueness}, \textit{Novelty}, \textit{Complete Rate}, and \textit{Scaffold Similarity} to measure models' capability in scaffold hopping. Detailed calculation of 
these metrics as \revise{follows:}
\begin{itemize}[leftmargin=*]
    \item \textbf{Validity} is defined as the fraction of generated molecules that can be successfully sanitized. 
    \item \textbf{Uniqueness} measures the proportion of unique molecules among the generated molecules.
    \item \textbf{Novelty} measures the fraction of generated molecules that not presented in training set.
    \item \textbf{Complete Rate} measures the proportion of completed molecules within the generated results.
    \item \textbf{Scaffold Similarity} Following \citet{polykovskiy2020molecular}, Bemis–Murcko scaffolds are extracted using rdkit function \texttt{MurckoScaffold}. We count the occurrences of scaffolds in all generated and reference molecules, creating vectors $G$ and $R$, where each dimension represents the count of a specific scaffold. The scaffold similarity is calculated as the cosine similarity between vectors $G$ and $R$.
\end{itemize}

\paragraph{More Examples of Generated Results}

\begin{figure}[H]
\centering
% \raggedleft
% \flushleft
\includegraphics[width=0.8\textwidth]{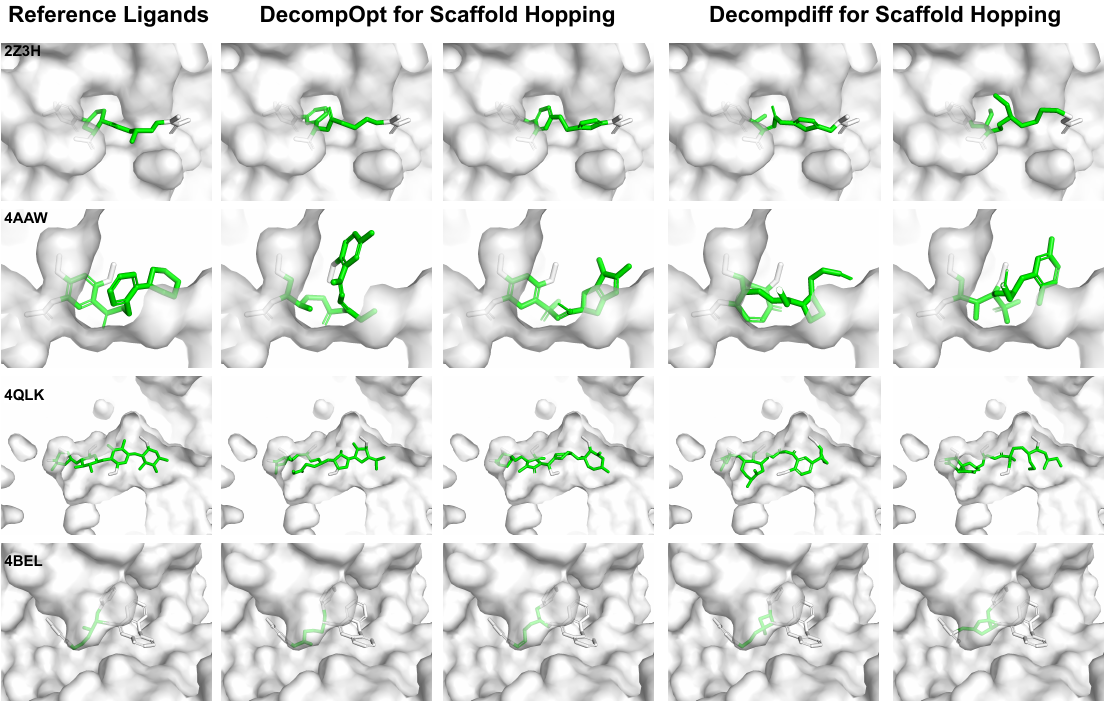}
  \caption{\revise{More examples of Scaffold Hopping results. The left column shows reference ligands, Scaffold Hopping results generated by \method are shown at the second and the third rows, and results generated by DecompDiff are shown at the fourth and the fifth rows. Scaffold are highlighted in green.}}{\label{fig:scaffold_hopping_additional}}
\end{figure}

\revise{For scaffold hopping, we provide more visualization of ligands generated by \method and DecompDiff on protein 2Z3H, 4AVW, 4QLK, and 4BEL, which are shown in \cref{fig:scaffold_hopping_additional}.}

\revise{\paragraph{Changes in Molecular Properties After Scaffold Hopping} Scaffold hopping aims at finding scaffold structures that can connect existing functional groups without disrupting their interactions with the target protein. The main purpose of this is to find novel scaffolds which are not protected by existing patents while maintaining comparable properties as the original molecule. Therefore, we did not implement property optimization mechanisms in scaffold hopping tasks and solely focusing on designing scaffolds that can connect existing arms. We provide the property comparison before and after scaffold hopping in \cref{tab:scaffold_hopping_properties}. As the result shows, the properties of the ligands remain relatively consistent before and after the process of scaffold hopping.}

\begin{table}[h]
    \centering
    \caption{\revise{Summary of properties of reference molecules and molecules generated through scaffold hopping using \method. ($\uparrow$) / ($\downarrow$) denotes a larger / smaller number is better.}}
    \begin{adjustbox}{width=1\textwidth}
    \renewcommand{\arraystretch}{1.2}
    \begin{tabular}{c|cc|cc|cc|cc|cc}
    \toprule
    % \diagbox{Model}{Metric} 
    \multicolumn{1}{c|}{\multirow{2}{*}{Methods}} & \multicolumn{2}{c|}{Vina Score ($\downarrow$)} & \multicolumn{2}{c|}{Vina Min ($\downarrow$)} & \multicolumn{2}{c|}{Vina Dock ($\downarrow$)} & \multicolumn{2}{c|}{QED ($\uparrow$)}   & \multicolumn{2}{c}{SA ($\uparrow$)} \\
    \multicolumn{1}{c|}{} & Avg. & Med. & Avg. & Med. & Avg. & Med. & Avg. & Med. & Avg. & Med. \\
    
    \toprule
    
    \multicolumn{1}{c|}{Reference}   & -6.36 & -6.46 & -6.71 & -6.49 & -7.45 & -7.26 & 0.48 & 0.47 & 0.73 & 0.74 \\

    Scaffold Hopping by \method & -5.89 & -6.13 & -6.46  & -6.28 &  -7.28 & -7.48 & 0.49 & 0.48 & 0.71 & 0.69 \\

    \bottomrule
    \end{tabular}
    \renewcommand{\arraystretch}{1}
    \end{adjustbox}\label{tab:scaffold_hopping_properties}
\end{table}

\end{document}